\documentclass[prd,nofootinbib,a4paper,showpacs,showkeys,preprintnumbers,twocolumn]{revtex4}

\usepackage{graphicx}
\usepackage{amsmath}
\usepackage{amssymb}
\usepackage{amsfonts}
\usepackage{latexsym}
\usepackage{mathrsfs}
\usepackage{color}
\usepackage{slashed}
\usepackage{feynmp}
\usepackage{pstricks,pst-text}
\usepackage{epsfig}
\usepackage{hyperref}
\usepackage{dsfont}

\providecommand{\higgs}{\phi}
\providecommand{\lepton}{{\ell}}
\providecommand{\leptdoublet}{{\ell}}
\providecommand{\majneutrino}{N}
\providecommand{\yu}{h}
\providecommand{\Ftilde}[4]{{\tilde{\mathcal{F}}_{#1 \leftrightarrow #2}^{#3 ; #4}}}
\providecommand{\F}[4]{{\mathcal{F}_{#1 \leftrightarrow #2}^{#3 ; #4}}}
\providecommand{\dpi}[4]{{d\Pi_{#1#2}^{#3#4}}}
\providecommand{\EffAmplitude}[2]{\Xi^{#1}_{#2}}
\providecommand{\EffAmpl}[3]{\Xi^{#1}_{#2 \leftrightarrow #3}}
\providecommand{\momk}{k}
\providecommand{\momp}{p}
\providecommand{\momq}{q}
\providecommand{\momlep}{\momp}
\providecommand{\momhig}{\momk}
\providecommand{\mommaj}{\momq}
\renewcommand{\vec}[1]{{\bf #1}}
\providecommand{\lorentzdd}[1]{\dif\Pi^4_{#1}}
\providecommand{\lorentzd}[2]{\dif\Pi^{#1}_{#2}}
\providecommand{\f}[2]{f^{#2}_{#1}}
\providecommand{\qstatff}[2]{(1-\f{#1}{#2})}
\providecommand{\dend}{{\, .}}
\providecommand{\kend}{{\, ,}}

\providecommand{\CP}{\textit{CP}}
\providecommand{\eqref}[1]{(\ref{#1})}
\providecommand{\fig}{Fig.\,}
\providecommand{\figs}{Figs.\,}

\providecommand{\sect}{Sec.\,}
\providecommand{\app}{Appendix\,}
\providecommand{\sign}{\mathrm{sign}}
\providecommand{\deltafour}[1]{\delta(#1)}
\providecommand{\dif}{\,d}
\providecommand{\tr}{{\rm tr}}
\providecommand{\ifeqthenelse}[4]{\edef\tempa{#1}\def\tempb{#2}\ifx\tempa\tempb {#3} \else {#4}\fi}
\providecommand{\pLept}[2]{S^{#1}_{#2}}
\providecommand{\pLeptcp}[2]{\bar{S}^{#1}_{#2}}
\providecommand{\pLeptmat}[2]{\hat{S}^{#1}_{#2}}
\providecommand{\pLeptsc}[2]{\mathbf{S}^{#1}_{#2}}
\providecommand{\pLeptEQP}[2]{\tilde{S}^{#1}_{#2}}
\providecommand{\sLept}[2]{\Sigma^{#1}_{#2}}
\providecommand{\sLeptcp}[2]{\bar{\Sigma}^{#1}_{#2}}
\providecommand{\sLeptmat}[2]{\hat{\Sigma}^{#1}_{#2}}
\providecommand{\pMaj}[2]{\mathscr{S}^{#1}_{#2}}

\providecommand{\pMajmat}[2]{\hat{\mathscr{S}}^{#1}_{#2}}

\providecommand{\pMajdiagmat}[2]{\hat{\mathcal{S}}^{#1}_{#2}}
\providecommand{\pMajdiagsc}[2]{\boldsymbol{\mathcal{S}}^{#1}_{#2}}

\providecommand{\pMajEQP}[2]{\tilde{\mathcal{S}}^{#1}_{#2}}
\providecommand{\pMajEQPsc}[2]{\tilde{\boldsymbol{\mathcal{S}}}^{#1}_{#2}}
\providecommand{\pMajEQPmat}[1]{\hat{\tilde{\mathcal{S}}}_{#1}}
\providecommand{\sMaj}[2]{\Pi^{#1}_{#2}}
\providecommand{\sMajmat}[2]{\hat{\Pi}^{#1}_{#2}}
\providecommand{\MatTheta}[2]{\Theta^{#1}_{#2}}
\providecommand{\MatThetacp}[2]{\bar{\Theta}^{#1}_{#2}}
\providecommand{\HatMatTheta}[2]{\hat{\Theta}^{#1}_{#2}}
\providecommand{\pHiggs}[2]{\Delta^{#1}_{#2}}
\providecommand{\pHiggsEQP}[2]{\tilde{\Delta}^{#1}_{#2}}
\providecommand{\sHiggs}[2]{\Omega^{#1}_{#2}}
\providecommand{\yuqSqu}{|\yuq|^2}
\providecommand{\topq}{t}
\providecommand{\pTopq}[2]{{S_{\topq}}{}^{#1}_{#2}}
\providecommand{\pTopqsc}[2]{{\mathbf{S}_{\topq}}{}^{#1}_{#2}}
\providecommand{\pTopqEQP}[2]{{\tilde{S}_{\topq}}{}^{#1}_{#2}}
\providecommand{\momtop}{{p_\topq}}
\providecommand{\momtopslash}{\slashed{p}_\topq}
\providecommand{\Q}{Q}
\providecommand{\pQ}[2]{{S_{\Q}}{}^{#1}_{#2}}
\providecommand{\pQsc}[2]{{\mathbf{S}_{\Q}}{}^{#1}_{#2}}
\providecommand{\pQEQP}[2]{{\tilde{S}_{\Q}}{}^{#1}_{#2}}
\providecommand{\momQ}{{p_\Q}}
\providecommand{\momQslash}{\slashed{p}_\Q}

\providecommand{\yuq}{{\lambda}}
\renewcommand{\Im}{{\rm Im}}

\begin{document}
\preprint{MPP-2013-29}

\title{Systematic approach to $\Delta L=1$ processes in thermal leptogenesis}

\author{T. Frossard$^{a}$}
\email[\,]{tibor.frossard@mpi-hd.mpg.de}    

\author{A. Kartavtsev$^{b}$}
\email[\,]{alexander.kartavtsev@mpp.mpg.de}

\author{D. Mitrouskas$^{c}$}
\email[\,]{david.mitrouskas@math.lmu.de}

\affiliation{%
$^{a}$Max-Planck-Institut f\"ur Kernphysik, Saupfercheckweg 1, 69117 Heidelberg, Germany\\
$^{b}$Max-Planck-Institut f\"ur Physik, F\"ohringer Ring 6, 80805 M\"unchen, Germany\\
$^{c}$LMU M\"unchen, Mathematisches Institut, Theresienstr. 39, 80333 M\"unchen, Germany}

\begin{abstract}
In this work we study the contribution to leptogenesis from $\Delta L=1$ decay and scattering 
processes mediated by the Higgs with quarks in the initial and final states using the formalism 
of non-equilibrium quantum field theory. Starting from fundamental equations for correlators of 
the quantum fields we derive quantum-corrected Boltzmann and rate equations for the total lepton 
asymmetry improved in that they include quantum-statistical effects and medium corrections to 
the quasiparticle properties. To compute the collision term we take into account  one- and 
two-loop contributions to the lepton self-energy and use the extended quasiparticle approximation 
for the Higgs two-point function. The resulting \CP-violating and washout reaction densities 
are numerically compared to the conventional ones.
\end{abstract}

\maketitle

% =============================================================================
\section{\label{introduction}Introduction}

The Standard Model (SM) of particle physics \cite{Weinberg:1967tq,Glashow:1961tr,Salam:1968rm} 
has successfully passed the numerous experimental tests performed so far. The recent observation 
of the Higgs particle \cite{Higgs:1964ia} at the LHC \cite{CMS:2012,ATLAS:2012} also seems 
to confirm the mechanism of spontaneous symmetry breaking, which is responsible for masses of 
the known gauge bosons and fermions. On the other hand,  we know that the SM is not 
complete. Firstly, it does not provide a viable dark matter candidate. Secondly, it predicts 
that the active neutrinos are strictly massless, which contradicts the results of neutrino 
oscillation experiments. A simple yet elegant way to generate
small but nonzero neutrino masses is to add three right-handed 
Majorana neutrinos to the model:
\begin{align} 
	\label{lagrangian}
	\mathscr{L}=\mathscr{L}_{SM}&+{\textstyle\frac12}\bar \majneutrino_i
	\bigl(i\slashed{\partial} -  M_i\bigr)\majneutrino_i\nonumber\\
	&- \yu_{\alpha i}\bar \leptdoublet_\alpha {\tilde \higgs}  P_R \majneutrino_i
	-\yu^\dagger_{i\alpha} \bar \majneutrino_i {\tilde \higgs}^\dagger   P_L \leptdoublet_\alpha\kend
\end{align}
where $\majneutrino_i=\majneutrino^c_i$ are the heavy Majorana fields, $\leptdoublet_\alpha$ are the 
lepton doublets and $\tilde \higgs\equiv i\sigma_2 \higgs^*$ is the conjugate of the Higgs doublet. 
After the electroweak symmetry breaking the active neutrinos receive naturally small masses through 
the type-I seesaw mechanism. This scenario has even more far-reaching consequences  as it can 
explain another beyond-the-SM observation, the baryon asymmetry of the universe.
The Majorana mass term in \eqref{lagrangian} violates lepton number. In the early 
Universe a decay of the Majorana neutrino into a lepton-Higgs pair increases the total lepton number of 
the Universe by one unit, and a decay into the corresponding antiparticles decreases the total lepton 
number by one unit. If there is \CP-violation then, on average, the number of leptons produced in those 
decays is not equal to the number of antileptons and a net lepton asymmetry is produced. It is also known 
that whereas the difference of the lepton and baryon numbers is conserved in the Standard Model, any 
other their linear combination is not \cite{Kuzmin:1985mm}. This implies that the lepton asymmetry 
produced by the Majorana neutrinos is partially converted to the baryon asymmetry \cite{Fukugita:1986hr}. 
This mechanism, which is referred to as baryogenesis via leptogenesis, naturally explains the observed 
baryon asymmetry of the Universe. For a more detailed review of leptogenesis see e.g. 
\cite{Buchmuller:2004nz,Davidson:pr2008,Drewes:2013gca}.

The state-of-the-art analysis of the asymmetry generation uses Boltzmann equations with the decay 
and scattering amplitudes calculated in vacuum. Their applicability in the hot and expanding early 
universe is questionable and can be cross-checked using a first-principle approach based on the use of 
non-equilibrium quantum field theory. One of the most important processes for the generation of the asymmetry 
is the decay of the Majorana neutrino. Thermal effects enhancing \CP-violation in the decay
have been studied in \cite{Garny:2009rv,Garny:2009qn,Beneke:2010wd,Garbrecht:2010sz,Garny:2010nj}.
The role of the flavor effects has been addressed in \cite{Beneke:2010dz}. A first-principle 
analysis of the asymmetry generation in the very interesting regime of resonant leptogenesis 
has been presented in \cite{Garny:2011hg} and \cite{Garbrecht:2011aw}. The effect of 
next-to-leading order corrections from the gauge interactions of lepton and
Higgs doublets on the production and decay rate of right-handed neutrinos at finite temperature
has been recently studied in \cite{Garbrecht:2013gd,Garbrecht:2013bia}. 

The asymmetry generated in the Majorana decay is partially washed out by the inverse decay and scattering processes. 
The latter can be classified into two categories. The first category includes $\Delta L=2$ 
scattering processes mediated by the Majorana neutrinos. A first-principle analysis of such processes 
free of the notorious double-counting problem has been presented in \cite{Frossard:2012pc}. The second 
category includes $\Delta L=1$ decay and scattering processes mediated by the Higgs. The latter processes 
are also known to play an important role in the asymmetry generation and are addressed in the present paper. 

The outline of the paper is as follows. In \sect\ref{CanonicalApproach} we briefly review the 
canonical approach to the analysis of the $\Delta L=1$ processes and derive the corresponding 
amplitudes and reduced cross-sections. In \sect\ref{NEQFTapproach} we derive quantum-generalized 
Boltzmann equations for the lepton asymmetry, calculate the effective amplitudes of the Higgs-mediated 
scattering processes and compare them with the canonical ones. The obtained Boltzmann equations are 
used in \sect\ref{RateEquations} to derive a simple system of rate equations for the total lepton 
asymmetry. In section \sect\ref{Numerics} we present a numerical comparison of the corresponding reaction 
densities with the ones obtained using the canonical approach. A summary of the results is presented 
in \sect\ref{Summary}.

\section{\label{CanonicalApproach}Conventional approach}

In the scenario of thermal leptogenesis  lepton asymmetry is generated in the lepton number 
and \CP-violating decay of the heavy Majorana neutrinos.
\begin{figure}[h!]
\includegraphics[width=0.95\columnwidth]{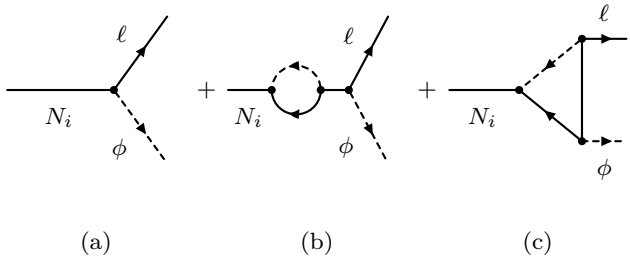}
\caption{\label{treevertexself}Tree-level, one-loop self-energy and one-loop vertex 
contributions to the decay of the heavy Majorana neutrino.}
\end{figure}
The corresponding \CP-violating parameters receive contributions from the interference of the 
tree-level amplitude with the vertex \cite{Fukugita:1986hr,Garny:2009rv} and self-energy 
\cite{Flanz:1994yx,Covi:1996wh,PhysRevD.56.5431,Pilaftsis:2003gt,Garny:2009qn} 
amplitudes, see \fig\ref{treevertexself}.  The contribution of the loop diagrams 
can be accounted for by effective Yukawa couplings \cite{Pilaftsis:2003gt}. If 
thermal masses of the SM particles are negligible, they are given by:
\begin{subequations}
\label{effective couplings}
\begin{align}
h_{+,\alpha i} &\equiv h_{\alpha i} -i h_{\alpha j} (h^\dagger h)^*_{ji}\, g_{ij}\kend\\
h_{-,\alpha i} &\equiv h^*_{\alpha i} -i h^*_{\alpha j} (h^\dagger h)_{ji}\, g_{ij}\kend
\end{align}
\end{subequations}
where the loop-function $g_{ij}$ is defined as 
\begin{align}
\label{LoopFunction}
g_{ij}\equiv &\frac{1}{16\pi} \frac{M_i M_j}{M^2_i-M^2_j}\nonumber\\
+ &\frac{1}{16\pi}\frac{M_j}{M_i}\biggl[1-\biggl(1+\frac{M_j^2}{M_i^2}\biggr)
\ln \biggl(1+\frac{M_i^2}{M_j^2}\biggr)\biggr]\dend
\end{align}
Note that this expression is valid only for on-shell final states. The first term in  
\eqref{LoopFunction} is related to the self-energy and 
the second term to the vertex contribution. This expression is applicable for 
a mildly or strongly hierarchical mass spectrum of the Majorana neutrinos. 
\begin{figure}[h!]
\includegraphics[width=0.27\columnwidth]{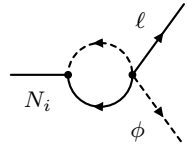}
\caption{\label{hierarchicalloop} Effective one-loop diagram for the self-energy and 
vertex contributions to the decay of the lightest Majorana neutrino for a strongly 
hierarchical mass spectrum.}
\end{figure}
In both cases most of the asymmetry is typically generated by the lightest Majorana 
neutrino, whereas the asymmetry generated by the heavier ones is almost completely
washed out. 
For a strongly hierarchical mass spectrum, $M_i\ll M_j$, the intermediate
Majorana line in \figs\ref{treevertexself}.b and \ref{treevertexself}.c 
contracts to a point, see \fig\ref{hierarchicalloop}, and the structure of 
the self-energy and vertex contributions is the same. In this limit:
\begin{align}
\label{LoopFunctionHierarchical}
g_{ij}\approx -\frac{3}{32\pi}\frac{M_i}{M_j}\dend 
\end{align}
Note that in this approximation the loop integral leading to \eqref{LoopFunctionHierarchical}
depends only on the momentum of the initial state and is independent of the momenta 
of the final states. This implies in particular that this expression can also be used for 
off-shell final states.

\begin{figure}[h!]
	\includegraphics[width=0.95\columnwidth]{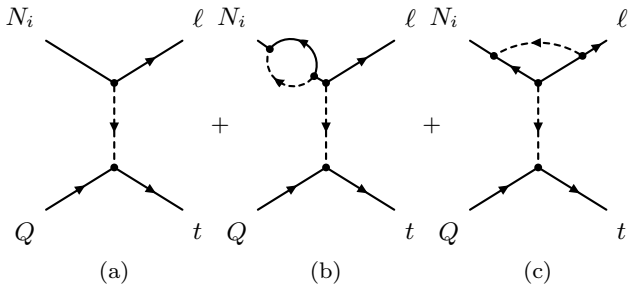}
	\caption{\label{fig:NQlt} Tree-level, self-energy and vertex contributions to the 
	scattering processes $\majneutrino_i \Q\rightarrow \lepton \topq$. Similar diagrams for the scattering
	process $\majneutrino_i \bar \topq \rightarrow \lepton \bar \Q$ are obtained by replacing $\Q$ with 
	$\bar \topq$ and $\topq$ with $\bar \Q$ as well as inverting the direction of the arrows.}
\end{figure}

\begin{figure}[h!]
	\includegraphics[width=0.95\columnwidth]{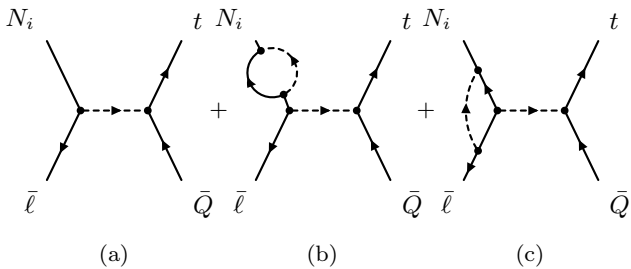}
	\caption{\label{fig:NlQt} Tree-level, self-energy and vertex contributions to the scattering 
	processes $\majneutrino_i \bar \lepton \rightarrow \bar \Q \topq$.}
\end{figure}

Using the effective couplings \eqref{effective couplings} we find for the decay amplitudes 
(squared) \cite{Pilaftsis:2003gt,Frossard:2012pc}:
\begin{subequations}
\label{EffAmplMajDecay}
\begin{align}
\EffAmplitude{}{\majneutrino_i\rightarrow\lepton \higgs} 
& = g_\majneutrino g_w (h_+^\dagger h^{\vphantom{\dagger}}_+)_{ii} (\mommaj\momlep)\kend\\
\EffAmplitude{}{\majneutrino_i\rightarrow\bar\lepton\bar\higgs} 
& = g_\majneutrino g_w (h_-^\dagger h^{\vphantom{\dagger}}_-)_{ii} (\mommaj\momlep)\kend
\end{align}
\end{subequations}
where we have summed over flavors of the leptons in the final state as well as over 
the Majorana spin ($g_\majneutrino=2$) and the $SU(2)_L$ group ($g_w=2$) degrees of 
freedom. Here $\mommaj$ and $\momlep$ are momenta of the heavy neutrino and
lepton, respectively. The decay amplitudes \eqref{EffAmplMajDecay} can be traded for 
the total decay amplitude and \CP-violating parameter:
\begin{subequations}
	\label{AmplsqAndEpsDef}
	\begin{align}
		\EffAmplitude{}{\majneutrino_i} & \equiv 
		\EffAmplitude{}{\majneutrino_i \rightarrow \lepton\higgs}+
		\EffAmplitude{}{\majneutrino_i \rightarrow \bar\lepton\bar\higgs}\kend\\
		\label{EpsilonUnintegrated}
		\epsilon_i & \equiv
		\frac{\EffAmplitude{}{\majneutrino_i \rightarrow \lepton\higgs}-
		\EffAmplitude{}{\majneutrino_i \rightarrow \bar\lepton\bar\higgs}}{
		\EffAmplitude{}{\majneutrino_i \rightarrow \lepton\higgs}+
		\EffAmplitude{}{\majneutrino_i \rightarrow \bar\lepton\bar\higgs}}\dend
	\end{align}
\end{subequations}
Combining \eqref{EffAmplMajDecay} and \eqref{AmplsqAndEpsDef} we then find for the 
(unflavored) \CP-violating parameter: 
\begin{align}
\label{epsilon vacuum}
\epsilon^{vac}_i\approx \frac{\Im(h^\dagger h)^2_{ij}}{(h^\dagger h)_{ii}}\times 2g_{ij}\kend
\quad j\neq i \dend
\end{align}
The asymmetry generated by the Majorana decay is partially washed out by the inverse 
decay and scattering processes violating lepton number. An important role is played 
by the $\Delta L=2$ scattering processes mediated by the heavy Majorana neutrinos 
\cite{Pilaftsis:2003gt,Plumacher:1997ru,Frossard:2012pc}. In addition, there are 
$\Delta L=1$ scattering process mediated by the Higgs doublet with quarks (or the 
gauge bosons) in the initial and final states \cite{Pilaftsis:2003gt,Plumacher:1997ru}, 
see  \fig\ref{fig:NQlt} and \fig\ref{fig:NlQt}. The Higgs coupling to the top is considerably 
larger than to the other quarks of the three generations. For this reason we do not consider 
the latter here. The corresponding Lagrangian reads:
\begin{align}
\label{toplagrangian}
\mathscr{L}_{SM}\supset -\yuq  \bar \Q \tilde \higgs P_R \topq 
-\yuq^* \bar \topq P_L {\tilde \higgs}^\dagger \Q\kend  
\end{align}
where $\Q$ and $\topq$ are the $SU(2)_L$ doublet and singlet of the third quark generation. 
The $\Delta L=1$ processes are also \CP-violating. The \CP-violation is generated by the same self-energy and 
vertex diagrams. Strictly speaking, since the Higgs is no longer on-shell the effective couplings 
\eqref{effective couplings} are not applicable in this case. 
On the other hand, for a strongly hierarchical   mass spectrum the intermediate Majorana lines in 
\fig\ref{fig:NQlt} and \fig\ref{fig:NlQt} again contract  to a point and the momenta of the Higgs and 
lepton play no role. In other words, for a strongly hierarchical mass spectrum we still can use the 
effective couplings \eqref{effective couplings} supplemented with \eqref{LoopFunctionHierarchical} 
to calculate the \CP-violating scattering amplitudes.

Summing over flavors and colors of the quarks and leptons in the initial and final states
as well as over the corresponding $SU(2)_L$ and spin degrees of freedom we find for the
amplitude of $\majneutrino_i\Q\rightarrow\lepton \topq$ scattering:
\begin{align}
\label{NQLtAmplitude}
\EffAmplitude{}{\majneutrino_i\Q\rightarrow\lepton \topq} = 
\EffAmplitude{}{\majneutrino_i\rightarrow\lepton \higgs} \times
\pHiggs{2}{T}(\momlep-\mommaj)\times
\EffAmplitude{}{\higgs \Q\rightarrow \topq}\kend
\end{align}
where $\pHiggs{}{T}(\momhig)\approx 1/(\momhig^2-m_\higgs^2)$ is the Feynman (or time-ordered)
propagator\footnote{In the kinematic region of interest the decay width term in the Feynman 
propagator of the Higgs plays no role and can be neglected.} of the intermediate  Higgs and 
we have defined 
\begin{align}
\label{VacuumEffDecAmpl}
\EffAmplitude{}{\higgs \Q\rightarrow \topq} & =  
2 g_s \yuqSqu (\momQ \momtop) \dend 
\end{align}
Here $g_s=3$ is the $SU(3)_C$ factor, and $p_\topq$ and $p_\Q$ are the momenta of 
the singlet and the doublet respectively. For the charge-con\-ju\-ga\-te process we find 
an expression similar to \eqref{NQLtAmplitude}. As can be inferred from \eqref{VacuumEffDecAmpl} 
in this work we neglect \CP-violation in the quark sector, which is known to be small. Defining 
\CP-violating parameter in scattering as 
\begin{align}
\label{defCPSE}
\epsilon_{X\rightarrow Y} \equiv \frac{\EffAmplitude{}{X\rightarrow Y}-\EffAmplitude{}{\bar X\rightarrow \bar Y}}{
\EffAmplitude{}{X\rightarrow Y}+\EffAmplitude{}{\bar X\rightarrow \bar Y}}\kend
\end{align}
we then obtain for $\epsilon_{\majneutrino_i\Q\rightarrow \lepton \topq}$ the same expression 
as for the Majorana decay, see \eqref{epsilon vacuum}. In the same approximation amplitude and \CP-violating 
parameter for $\majneutrino_i\bar\topq \rightarrow \lepton \bar\Q$ scattering coincide 
with those for $\majneutrino_i\Q\rightarrow \lepton \topq$ process.
Proceeding in a similar way we find for the scattering amplitude of the 
$\majneutrino_i \bar \lepton \rightarrow \bar \Q \topq$ process: 
\begin{align}
\label{NLQtAmplitude} 
\EffAmplitude{}{\majneutrino_i \bar \lepton \rightarrow \bar \Q \topq} = 
\EffAmplitude{}{\majneutrino_i\bar\lepton\rightarrow\higgs} \times
\pHiggs{2}{T}(\momlep+\mommaj)\times
\EffAmplitude{}{\higgs\rightarrow \bar\Q\topq}\kend
\end{align}
where $\EffAmplitude{}{\higgs\rightarrow \bar\Q\topq}=\EffAmplitude{}{\higgs\Q\rightarrow \topq}$
because we neglect \CP-vi\-o\-la\-tion in the quark sector. Furthermore, for a strongly hierarchical mass spectrum 
$\EffAmplitude{}{\majneutrino_i\bar\lepton\rightarrow\higgs}=\EffAmplitude{}{\majneutrino_i\rightarrow\lepton\higgs}$.
The resulting expression for the \CP-violating parameter then coincides with \eqref{epsilon vacuum}.  

\begin{figure}[h!]
	\includegraphics[width=0.95\columnwidth]{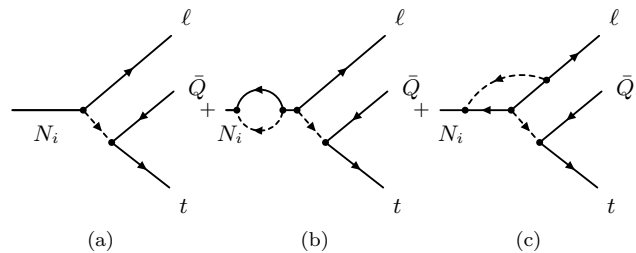}
	\caption{\label{fig:Ampl_N_lQt} Tree-level, self-energy and vertex contributions to the 
	amplitude of the three-body decay processes $\majneutrino_i\rightarrow \lepton \bar \Q \topq$.}
\end{figure}

If the lepton and both quarks are in the final state  then instead of a scattering process we deal 
with a three-body Majorana decay, see \fig\ref{fig:Ampl_N_lQt}.
In complete analogy with the scattering processes we can write its amplitude in the form:
\begin{align}
\label{NtoLQtAmplitude} 
\EffAmplitude{}{\majneutrino_i \rightarrow \lepton  \bar \Q \topq} = 
\EffAmplitude{}{\majneutrino_i\rightarrow\lepton\higgs} \times
\pHiggs{2}{T}(\momQ+\momtop)\times
\EffAmplitude{}{\higgs\rightarrow \bar\Q\topq}\dend
\end{align}
Evidently, \CP-violating parameter for this process coincides with that for the 
two-body Majorana decay.

To compute the generated lepton asymmetry, the conventional approach uses the 
generalized Boltzmann equation for the total lepton abundance, $Y_L\equiv n_L/s$, with
$s$ being the comoving en\-tro\-py density \cite{Kolb:1990vq}. In the Friedmann-Robertson-Walker (FRW) universe 
the contribution of the Higgs-mediated processes to the right-hand side of the 
the Boltzmann equation  simplifies to:
\begin{align}
 	\label{ConvBltzmnEq}
	 \frac{s\cal{H}}{z}&\frac{d Y_L}{dz}=\ldots\nonumber\\
	 &-\sum_i\int \dpi{\majneutrino_i \lepton}{\Q\topq}{\mommaj\momlep}{\momQ\momtop}
	 (2\pi)^4\delta(\mommaj+\momlep-\momQ-\momtop)
	 \nonumber\\
	 &\times \bigl[\EffAmplitude{}{\majneutrino_i\lepton\rightarrow\Q \bar\topq}
	 \f{\majneutrino_i}{}\f{\lepton}{}\qstatff{\Q}{}\qstatff{\bar\topq}{}-{\rm inverse}\bigr]	 
	 \nonumber\\
	 &+\sum_i\int \dpi{\majneutrino_i \Q}{\lepton\topq}{\mommaj\momQ}{\momlep\momtop}
	 (2\pi)^4\delta(\mommaj+\momQ-\momlep-\momtop)
	 \nonumber\\
	 &\times \bigl[ \EffAmplitude{}{\majneutrino_i\Q\rightarrow\lepton\topq}
	 \f{\majneutrino_i}{}\f{\Q}{}\qstatff{\lepton}{}\qstatff{\topq}{}-{\rm inverse}\bigr]	 
	 \nonumber\\
	 &+\sum_i\int \dpi{\majneutrino_i \bar\topq}{\lepton\bar\Q}{\mommaj\momtop}{\momlep\momQ}
	 (2\pi)^4\delta(\mommaj+\momtop-\momlep-\momQ)
	 \nonumber\\
	 &\times \bigl[ \EffAmplitude{}{\majneutrino_i\bar\topq\rightarrow\lepton\bar\Q}
	 \f{\majneutrino_i}{}\f{\bar\topq}{}\qstatff{\lepton}{}\qstatff{\bar\Q}{}-{\rm inverse}\bigr]	 
	 \nonumber\\
	 &+\sum_i\int \dpi{\majneutrino_i}{\lepton\bar\Q \topq}{\mommaj}{\momlep\momQ\momtop}
	 (2\pi)^4\delta(\mommaj-\momlep-\momQ-\momtop)
	 \nonumber\\
	 &\times \bigl[ \EffAmplitude{}{\majneutrino_i\rightarrow\lepton\bar\Q\topq}
	 \f{\majneutrino_i}{}\qstatff{\lepton}{}\qstatff{\bar\Q}{}\qstatff{\topq}{}-{\rm inverse}\bigr]
	 \nonumber\\
	 & - {\rm CP\,\, conjugate\,\,processes}.
\end{align} 
where we have introduced the dimensionless inverse temperature $z=M_1/T$, the Hubble 
rate ${\cal H}=H{|_{T=M_1}}$, and $\dpi{ab \dotso}{ij \dotso}{p_a p_b \dotso}{p_i p_j \dotso}$ 
stands for the product of the invariant phase space elements, $\lorentzd{p}{a}\equiv d^3p/[(2\pi)^3 2E_p]$.
Note that to ensure vanishing of the asymmetry in thermal equilibrium one should also include \CP-violating 
$2\leftrightarrow 3$ processes \cite{Davidson:pr2008}. Since there is no need for that in the 
non-equilibrium QFT approach we will not consider these processes here.

\section{\label{NEQFTapproach}Non-equilibrium QFT approach}

The formalism of non-equilibrium quantum field theory provides a powerful tool for the description 
of out-of-equilibrium quantum fields and is therefore well suited for the analysis of leptogenesis. 
In this section we briefly review results obtained in \cite{Frossard:2012pc} and introduce notation 
that will be used in the rest of the paper. As has been argued in \cite{Frossard:2012pc}, the equation 
of motion for the lepton asymmetry can be derived by considering the divergence of the lepton current. 
In the FRW Universe $j_L^\mu=(n_L,\vec{0})$ and therefore it is related to the total lepton abundance by:
\begin{align}
	{\cal D}_{\mu} j_L^\mu  = \frac{s{\cal H}}{z}\frac{d Y_L}{dz}\dend
\end{align}
Using the formalism of non-equilibrium quantum field theory one can express it through propagators 
and self-energies of leptons. After a transformation to the Wig\-ner space we obtain \cite{Frossard:2012pc}:
\begin{align}
\label{lepcurwig}
{\cal D}_\mu j_L^\mu (t,p)=g_w\int  \! \lorentzdd{\momlep} \, \tr\bigl[ & \sLeptmat{}{<}(t,p) \pLeptmat{}{>} (t,p)
\nonumber \\ &- \sLeptmat{}{>}(t,p) \pLeptmat{}{<}(t,p) \bigr] \kend
\end{align}
where $\lorentzdd{\momlep}\equiv {d^4\momlep}/{(2\pi)^4}$  and the hats denote matrices 
in flavor space. In \eqref{lepcurwig} we have taken into account that the $SU(2)_L$ symmetry 
is unbroken at the epoch of leptogenesis. As a consequence, the $SU(2)_L$ structure of the 
propagator is trivial, $\pLept{\alpha\beta}{ab}=\delta_{ab}\pLept{\alpha\beta}{}$, and 
summation over the $SU(2)_L$ components simply results in the overall factor $g_w=2$. Furthermore, 
in this work we restrict ourselves to the analysis of unflavored leptogenesis. Therefore, the 
lepton propagator can be approximated by $\pLept{\alpha \beta }{}=\delta^{\alpha\beta} \pLept{}{}$.
Similar relation also holds for the lepton self-energy. Then the equation for the divergence 
of the lepton current takes the form:
\begin{align}
	\label{MasterEquationWigner}
	{\cal D}_\mu j_L^\mu & =  g_w \int\limits_0^\infty\frac{dp^0}{2\pi} \int \frac{d^3\momlep}{(2\pi)^3}\\
	&\times \tr \bigl[ \bigl(\sLept{}{<}\pLept{}{>} - \sLept{}{>}\pLept{}{<}\bigr)
	-\bigl(\sLeptcp{}{<} \pLeptcp{}{>}-\sLeptcp{}{>} \pLeptcp{}{<} \bigr)\bigr]\kend\nonumber
\end{align}
where $\sLept{}{}\equiv \sLept{\alpha\alpha}{}$ and we have suppressed the argument $(t,p)$ of the 
two-point functions.  Note that the trace in \eqref{MasterEquationWigner}
acts now in spinor space only. 
To convert the integration over positive and negative frequencies into the integration over 
positive frequencies only we have introduced in \eqref{MasterEquationWigner} \CP-conjugate 
two-point functions and self-energies which are denoted by the bar. According to the extended 
quasiparticle approximation (eQP) \citep{CondMatPhys2006_9_473,PhysRevC.48.1034,JPhys2006_35_110} 
the Wigthmann propagators can be split into on- and off-shell parts:
\begin{align}
 \label{plepeqp}
\pLept{}{\gtrless}=\pLeptEQP{}{\gtrless}-{\textstyle\frac{1}{2}}\bigl(\pLept{}{R} \sLept{}{\gtrless} \pLept{}{R}
+ \pLept{}{A} \sLept{}{\gtrless} \pLept{}{A}\bigr) \dend
\end{align}
The off-shell parts of the lepton propagators exactly cancel out in the lepton current as they 
are lepton number conserving. On the other hand, as we will see later, the off-shell part of the 
Higgs two-point functions is crucial for a correct description  of the scattering processes. 
The on-shell part of the Wightman propagators is related to the eQP spectral function and 
one-particle distribution function $f_\lepton$ by the Kadanoff-Baym (KB) ansatz:
\begin{align}
 \label{eqpspectral}
\pLeptEQP{}{>} &=\left(1-f_\lepton  \right)\pLeptEQP{}{\rho}  
\kend \quad \pLeptEQP{}{<} =-f_\lepton\,  \pLeptEQP{}{\rho}  
 \kend
\end{align}
where
\begin{align}
\label{LeptonEQP}
 \pLeptEQP{}{\rho}=-\frac{1}{2}\pLept{}{R} \sLept{}{\rho}
 \pLept{}{R}\sLept{}{\rho}\pLept{}{A}\sLept{}{\rho}\pLept{}{A} \dend
\end{align}
In the limit of vanishing width the eQP spectral function $\pLeptEQP{}{\rho}$ approaches 
the Dirac delta-function \cite{Frossard:2012pc},
\begin{align}
\label{leprhotilde}
\pLeptEQP{}{\rho}  \approx (2\pi)\,\sign(\momlep^0) \delta(\momlep^2-m_\lepton^2) &
P_L \slashed{\momlep} P_R \nonumber\\
&\equiv \pLeptsc{}{\rho}(p) P_L \slashed{\momlep} P_R \kend
\end{align}
where we have extracted the `scalar' part $\pLeptsc{}{\rho}$ for notational convenience. 
In \eqref{leprhotilde} we have approximately taken the gauge interactions into account in 
the form of effective masses of the leptons. Note that we will not attempt
a fully consistent inclusion of the gauge interactions here. In the used approximation the 
spectral function is \CP-symmetric.  This implies that the spectral properties, in particular 
the masses, of the particles and antiparticles are the same.

To evaluate the right-hand side of  \eqref{MasterEquationWigner} we need to specify the form 
of the lepton self-energy. It can be obtained by functional differentiation of the 
2PI effective action with respect to the lepton propagator. Loosely speaking, this means 
that the self-energies are obtained by cutting one line of the 2PI contributions to the 
effective action.
\begin{figure}[h!]
\includegraphics[width=0.7\columnwidth]{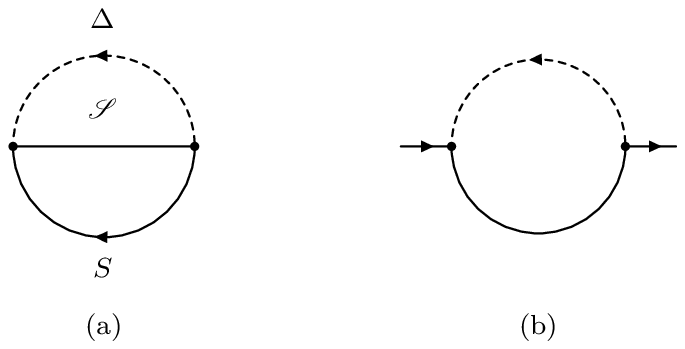}\\[2mm]
\includegraphics[width=0.7\columnwidth]{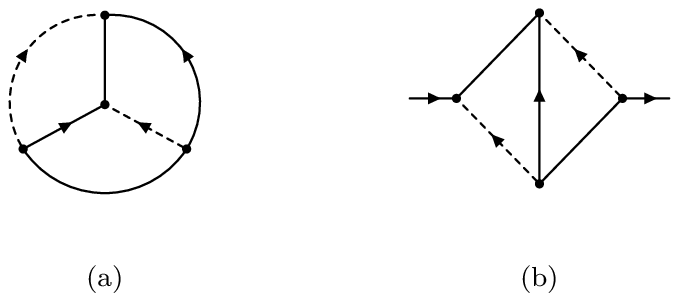}
\caption{\label{fig:2PI contributions}Two- and three-loop contributions to the 2PI 
effective action and the corresponding contributions to the lepton self-energy.}
\end{figure} 
The two- and three-loop contributions are presented in \fig\ref{fig:2PI contributions}(a)
and \fig\ref{fig:2PI contributions}(c).
The one-loop contribution to the lepton self-energy, see \fig\ref{fig:2PI contributions} (b), 
is given by \cite{Frossard:2012pc}:
\begin{align}
	\label{Sigma1WT}
	\sLept{(1)}{\gtrless}(t,\momlep) 
	= & -\int \lorentzdd{\mommaj \momhig }(2\pi)^4 
	\deltafour{\momlep+\momhig-\mommaj}\nonumber\\
	& \times 
	(\yu^\dagger\yu)_{ji} \, P_R \pMaj{ij}{\gtrless}(t,\mommaj) P_L \pHiggs{}{\lessgtr}(t,\momhig)\kend
\end{align}
where $\pMaj{}{}$ and $\pHiggs{}{}$ denote the Majorana and Higgs propagators respectively, and 
$\lorentzdd{\mommaj\momhig }\equiv \lorentzdd{ \mommaj}\lorentzdd{\momhig} $.
The expression for the two-loop contribution, see \fig\ref{fig:2PI contributions} (d), is 
rather lengthy. Here we will only need a part of it:
\begin{align}
	\label{SEDMT}
	&\sLept{(2)}{\gtrless}(t,\momlep) = \int \lorentzdd{\mommaj \momhig}\
	(2\pi)^4\delta(\momlep+\momhig-\mommaj) \\
	&\times\big[ (\yu^\dagger \yu)_{in}(\yu^\dagger \yu)_{jm}
	\varLambda_{mn}(t,q,k)P_L C \pMaj{ij}{\gtrless}(t,\mommaj) 
	P_L\pHiggs{}{\lessgtr}(t,\momhig)\nonumber\\
	&+ (\yu^\dagger \yu)_{ni} (\yu^\dagger \yu)_{mj}
	P_R \pMaj{ji}{\gtrless}(t,\mommaj) C P_R V_{nm}(t,\mommaj,\momhig)
	\pHiggs{}{\lessgtr}(t,\momhig)\big]\nonumber \kend
\end{align}
where we have introduced two functions containing loop corrections:
\begin{align}
	\label{LC1MT}
	&\varLambda_{mn}(t,\mommaj,\momhig) \equiv \int \lorentzdd{k_1 k_2 k_3}\nonumber\\
	& \times (2\pi)^4\delta(\mommaj+k_1+k_2)\ (2\pi)^4 \delta(\momhig+k_2-k_3)\nonumber\\
	& \times \bigl[ P_R \pMaj{mn}{R}(t,-k_3) C P_R \pLept{T}{F}(t,k_2)\pHiggs{}{A}(t,k_1)\nonumber\\
	& +\ P_R \pMaj{mn}{F}(t,-k_3) C P_R \pLept{T}{R}(t,k_2) \pHiggs{}{A}(t,k_1)\nonumber\\
	& +\ P_R \pMaj{mn}{R}(t,-k_3) C P_R \pLept{T}{A}(t,k_2) \pHiggs{}{F}(t,k_1) \bigr]\kend
\end{align}
and $V_{nm}(t,q,k)\equiv P\, \varLambda^\dagger_{nm}(t,q,k)\, P$ to shorten the notation.
Here $P=\gamma^0$ is the parity conjugation operator. The remaining terms of the two-loop 
self-energy can be found in \cite{Frossard:2012pc}. As has been demonstrated in 
the same reference, \CP-conjugates of the above self-energies can be obtained by 
replacing the Yukawa couplings by the complex conjugated ones and the propagators by 
the \CP-conjugated ones. 

Comparing  \eqref{Sigma1WT} and \eqref{SEDMT} we see that the two self-energies have a 
very similar structure. First, the integration is over momenta of the Higgs and Majorana 
neutrino and the delta-function contains the same combination of the momenta. Second, they 
both include one Wightman propagator of the Higgs field and one Wightman propagator of the 
Majorana field. These can be interpreted as cut-propagators which describe on-shell particles 
created from or absorbed by the plasma \cite{Carrington:2004tm}. The retarded and advanced 
propagators can be associated with the off-shell intermediate states. We therefore conclude 
that the two self-energies describe \CP-violating decay  of the heavy neutrino into a 
lepton-Higgs pair. Note that this interpretation only holds for the ``particle'' part of the 
eQP ansatz. The inclusion of the off-shell part of the Higgs Wightman propagator gives raise 
to the Higgs mediated scattering processes and three-body decay, see section \ref{HiggsMediatedScattering}.

To evaluate \eqref{Sigma1WT} and \eqref{SEDMT} we need to know the form of the Higgs and 
Majorana propagators. For the Higgs field we will adopt in this section a leading-order 
approximation:
\begin{align}
	\label{KBHiggs}
	\pHiggs{}{>}=(1+f_\higgs)\pHiggs{}{\rho}\kend\quad 
	\pHiggs{}{<}=f_\higgs\pHiggs{}{\rho}\kend
\end{align}
and a simple quasiparticle approximation for the spectral function,
\begin{align}
	\label{QPforHiggs}
	\pHiggs{}{\rho}(t,\momhig)=\,(2\pi)\,\sign(\momhig^0)\,\delta(\momhig^2-m_\higgs^2)\kend
\end{align}
where $m_\higgs$ is the effective thermal Higgs mass. Close to thermal equilibrium the full resummed 
Majorana propagator is given by \cite{Frossard:2012pc}: 
\begin{align}
	\label{eQPequ}
	\pMajmat{}{\gtrless} & =\HatMatTheta{}{R} \bigl[\,\pMajEQPmat{\gtrless}\nonumber\\
	&-\pMajdiagmat{}{R} \sMajmat{'}{\gtrless}\pMajdiagmat{}{A} 
	- {\textstyle\frac{1}{2}}\bigl(\pMajdiagmat{}{R} \sMajmat{d}{\gtrless}\pMajdiagmat{}{R}+
	\pMajdiagmat{}{A} \sMajmat{d}{\gtrless}\pMajdiagmat{}{A} \bigr) 
	\bigr]\HatMatTheta{}{A}\kend 
\end{align}
where $\sMajmat{d}{}$ and $\sMajmat{'}{}$ denote the diagonal and off-diagonal components 
of the Majorana self-energy respectively, $\pMajdiagmat{}{R}$ and $\pMajdiagmat{}{A}$ are given by
\begin{align}
	\label{diagRA}
	\pMajdiagmat{}{R(A)}=-\bigl( \slashed{\mommaj} -\hat{M} - \sMajmat{d}{R(A)}\bigr)^{-1}\kend
\end{align}
and we have introduced 
\begin{align}
      \HatMatTheta{}{R}\equiv \bigl(\mathds{1}+\pMajdiagmat{}{R}\sMajmat{'}{R} \bigr)^{-1}\kend\quad
      \HatMatTheta{}{A}\equiv \bigl(\mathds{1}+\sMajmat{'}{A} \pMajdiagmat{}{A} \bigr)^{-1}\kend
\end{align}
to shorten the notation. The first term in the square brackets of \eqref{eQPequ} describes (inverse)
decay of the Majorana neutrino, whereas the remaining three terms describe two-body scattering 
processes mediated by the Majorana neutrino. For the ``particle'' part of the eQP diagonal Wightman 
propagators of the Majorana neutrino one can use  the KB approximation:
\begin{align}
\label{eqpmaj}
\pMajEQP{nn}{>}=(1-f_{N_n}) \pMajEQP{nn}{\rho}\,,\quad 
\pMajEQP{nn}{<}=-f_{N_n} \pMajEQP{nn}{\rho}\kend
\end{align}
with the spectral function given by an expression identical to \eqref{LeptonEQP}.
Substituting \eqref{diagRA} we find in the limit of small decay width:
\begin{align}
\pMajEQP{nn}{\rho} =(2\pi)\,\sign(\mommaj^0)\delta(\mommaj^2-M^2_n) & (\slashed{\mommaj}+M_n)\nonumber\\
& \equiv \pMajEQPsc{nn}{\rho}(\slashed{\mommaj}+M_n) \dend
\end{align}
Inserting  \eqref{Sigma1WT} and \eqref{SEDMT} into 
the divergence of the lepton current \eqref{MasterEquationWigner} and integrating over 
the frequencies we then obtain an expression that strongly resembles the Boltzmann equation:
\begin{align}
	\label{asymLtree}
	\frac{s{\cal H}}{z}\frac{d Y_L}{dz} & = \sum_i \int 
	\dpi{\majneutrino_i}{\lepton\higgs}{q}{pk}\nonumber \\ 
 	& \times \bigl[\EffAmpl{}{\majneutrino_i}{\lepton\higgs} 
	\F{\majneutrino_i}{\lepton\higgs}{\mommaj}{\momlep \momhig} 
	- \EffAmpl{}{\majneutrino_i}{\bar\lepton\bar\higgs} 
	\F{\majneutrino_i }{\bar\lepton\bar\higgs}{\mommaj}{\momlep \momhig} \bigr]\kend
\end{align}
where we have introduced 
\begin{align}
	\label{eqn:definition of F}
	&\F{ab ..}{ij ..}{p_a p_b ..}{p_i p_j ..}  \equiv 
	(2\pi)^4 \delta(p_a+p_b+\ldots -p_i-p_j-\ldots)\nonumber\\
	&\hspace{10mm} \times\bigl[\f{a}{p_a} \f{b}{p_b} \ldots  (1\pm \f{i}{p_i})(1\pm \f{j}{p_j}) \ldots  \nonumber\\ 
	&\hspace{10mm} - \f{i}{p_i} \f{j}{p_j} \ldots (1\pm \f{a}{p_a})(1\pm \f{b}{p_b}) \ldots \bigr] \kend
\end{align}
with the plus (minus) sign corresponding to bosons (fermions). Note that $\F{ab ..}{ij ..}{p_a p_b ..}{p_i p_j ..}$ 
vanishes in equilibrium due to detailed balance. This implies that in accordance with the third 
Sakharov condition \cite{Sakharov:1967dj} no asymmetry is generated in equilibrium. In the 
Kadanoff-Baym formalism this result is obtained automatically and no need for the real 
intermediate state subtraction arises. 

The effective decay amplitudes $\EffAmplitude{}{}{}$ are given by a sum of the tree-level, one-loop self-energy and 
one-loop vertex contributions. The first two:
\begin{subequations}
	\label{MajoranaSelfEnAmplitudes}
	\begin{align}
		\EffAmpl{T}{\majneutrino_i}{\lepton\higgs}&+\EffAmpl{S}{\majneutrino_i}{\lepton\higgs}
		\equiv g_w{\textstyle\sum}_{mn}(h^\dagger h)_{mn}\nonumber\\
		&\times\tr[\MatTheta{ni}{R}(\mommaj)(\slashed{\mommaj}+M_i)
		\MatTheta{im}{A}(\mommaj)P_L\slashed{\momlep}P_R\,]\kend\\
		\EffAmpl{T}{\majneutrino_i}{\bar\lepton\bar\higgs}&
		+\EffAmpl{S}{\majneutrino_i}{\bar\lepton\bar\higgs}\equiv 
		g_w {\textstyle\sum}_{mn}(h^\dagger h)^*_{mn}\nonumber\\
		&\times\tr[\MatThetacp{ni}{R}(\mommaj)(\slashed{\mommaj}+M_i)
		\MatThetacp{im}{A}(\mommaj)P_L\slashed{\momlep}P_R\,]\kend
	\end{align}
\end{subequations}
emerge from the one-loop lepton self-energy \eqref{Sigma1WT}. The third one:
\begin{subequations}
\label{MajoranaVertexAmplitudes}
\begin{align}
	\EffAmpl{V}{N_i}{\lepton \higgs} \equiv & -g_w(h^{\dagger}h)_{ij}^2\ 
	M_i\,\tr\bigl[\varLambda_{jj}(q,k) C P_L\slashed pP_R\bigr]\nonumber\\
	&-g_w(h^{\dagger}h)_{ji}^{2}\,M_i\, \tr\bigr[ C V_{jj}(q,k)P_L\slashed p P_R\bigr]\kend\\
	\EffAmpl{V}{N_i}{\bar \lepton \bar \higgs} 
	\equiv & -g_w(h^{\dagger}h)_{ij}^2\ M_i\,
	\tr\bigr[ C V_{jj}(q,k)P_L\slashed p P_R\bigr]\nonumber\\
	&-g_w(h^{\dagger}h)_{ji}^{2}\,M_i\, \tr\bigl[\varLambda_{jj}(q,k)
	C P_L\slashed pP_R\bigr]\kend
\end{align}
\end{subequations}
is generated by the two-loop lepton self-energy \eqref{SEDMT}.
Substituting \eqref{MajoranaSelfEnAmplitudes} and \eqref{MajoranaVertexAmplitudes} into 
\eqref{AmplsqAndEpsDef} we find to leading order in the couplings that the total decay 
amplitude summed over the Majorana spin degrees of freedom is given by 
$\EffAmplitude{}{\majneutrino_i}=2g_N g_w (h^\dagger h)_{ii}(\momlep\mommaj)$.
The self-energy \CP-violating parameter reads \cite{Frossard:2012pc}:
\begin{align}
\label{CPparamSEdecay}
\epsilon_i^{S}&\approx 
-\sum \frac{\Im(\yu^\dagger \yu)_{ij}^2 }{(\yu^\dagger \yu)_{ii}(\yu^\dagger \yu)_{jj}}
\frac{M_i\Gamma_j}{M_j^2} \frac{\momlep L_S}{\mommaj\momlep} \cdot M_j^2 \pMajdiagsc{jj}{h}(\mommaj)
\kend
\end{align}
where the `scalar' part of the diagonal hermitian Majorana propagator is given by \cite{Frossard:2012pc}:
\begin{align}
\label{DiagonalHermitian}
\pMajdiagsc{jj}{h}(q)&\equiv {\textstyle\frac12}\bigl[\pMajdiagsc{jj}{R}(q)+\pMajdiagsc{jj}{A}(q)\bigr]\nonumber\\
&\approx -\frac{q^2-M_j^2}{(q^2-M_j^2)^2+(\Gamma_j/M_j\cdot qL_S)^2}\dend
\end{align}
It describes the intermediate Majorana neutrino line in \fig\ref{treevertexself}.b. Note that 
\eqref{CPparamSEdecay} has been obtained assuming a hierarchical mass spectrum of the heavy 
neutrinos and is not applicable for a quasidegenerate spectrum. For positive $\mommaj^0$ and 
$\mommaj^2$ the self-energy loop function $L_S$ is given by \cite{Frossard:2012pc}:
\begin{align}
	\label{eqn:Lrho}
	L^\mu_{S}=16\pi \int \lorentzd{\higgs \lepton}{\momhig_1 \momlep_1} & 
	(2\pi)^4 \deltafour{\mommaj-\momhig_1-\momlep_1}\,
	\momlep^\mu_1\nonumber\\
	&\times \bigl[1+\f{\higgs}{\momhig_1}-\f{\lepton}{\momlep_1}\bigr]\dend
\end{align}
Simplifying \eqref{MajoranaVertexAmplitudes} we find for the vertex \CP-violating parameter \cite{Frossard:2012pc}:
\begin{align}
\label{MajoranaCPVertex}
\epsilon^V_i &= -\frac12\sum \frac{\Im\,(h^\dagger h)_{ij}^2}{(h^\dagger h)_{ii}(h^\dagger h)_{jj}}
\frac{M_i\Gamma_j}{M_j^2}\frac{\momlep L_V}{\mommaj \momlep}\dend 
\end{align}
The vertex loop function is given by:
\begin{align}
	L^\mu_V&(q,p) = 16\pi\, M_j^2\int \lorentzdd{\mommaj_1 \momlep_1 \momhig_1} \\
	&\times (2\pi)^4 \delta(q+\momhig_1+\momlep_1) (2\pi)^4 \delta(\mommaj-\momlep+\momlep_1-\mommaj_1)\, 
	\momlep_1^\mu \nonumber\\ 
	&\times \bigl[ \pHiggs{}{\rho}(\momhig_1) \pLeptsc{}{F}(\momlep_1) \pMajdiagsc{jj}{h}(\mommaj_1) 
	+ \pHiggs{}{F}(\momhig_1) \pLeptsc{}{\rho}(\momlep_1) \pMajdiagsc{jj}{h}(\mommaj_1)\nonumber\\
	& - \pHiggs{}{h}(\momhig_1) \pLeptsc{}{\rho}(\momlep_1) \pMajdiagsc{jj}{F}(\mommaj_1) 
	+ \pHiggs{}{h}(\momhig_1) \pLeptsc{}{F}(\momlep_1) \pMajdiagsc{jj}{\rho}(\mommaj_1)
	\nonumber\\
	& + \pHiggs{}{\rho}(\momhig_1) \pLeptsc{}{h}(\momlep_1) \pMajdiagsc{jj}{F}(\mommaj_1) 
	+ \pHiggs{}{F}(\momhig_1) \pLeptsc{}{h}(\momlep_1) \pMajdiagsc{jj}{\rho}(\mommaj_1)\bigr]\nonumber\kend
\end{align}
where $\pMajdiagsc{}{F}=(\pMajdiagsc{}{>}+\pMajdiagsc{}{<})/2$ is  the `scalar' part of 
the corresponding statistical propagator of the heavy neutrino. For the lepton and Higgs fields 
the definitions are similar. The three lines in the square brackets in \eqref{MajoranaCPVertex} correspond to different 
cuts through two of the three internal lines of the vertex loop. The first line  corresponds to 
cutting the propagators of the Higgs and lepton and can be simplified to \cite{Garny:2010nj}:
\begin{align}
\label{cutlphi}
\momlep L^{\lepton \higgs}_V& (\mommaj,\momlep) = 16\pi \int  \lorentzd{\higgs \lepton}{\momhig_1 \momlep_1}
(2\pi)^4\delta(\mommaj-\momlep_1-\momhig_1) \nonumber\\ 
&\times (\momlep \momlep_1) \bigl[1+\f{\higgs}{\momhig_1}-\f{\lepton}{\momlep_1}\bigr]
\frac{M_j^2}{M_j^2-(\mommaj-\momlep_1-\momlep)^2} \dend 
\end{align}
The other two are cuts through the Majorana and lepton and the Majorana and Higgs lines 
respectively \cite{Garbrecht:2010sz}. For the second cut we obtain: 
\begin{align}
\label{cutNl}
\momlep L^{\majneutrino_j\lepton}_V (\mommaj,\momlep)& =16\pi\int  
\lorentzd{\majneutrino_j \lepton}{\mommaj_1\momlep_1} 
(2\pi)^4\delta(\mommaj-\momlep+\momlep_1-\mommaj_1) \nonumber\\ 
& \times(\momlep \momlep_1)\,\bigl[\f{\majneutrino_j}{\mommaj_1}-\f{\lepton}{\momlep_1}\bigr]
\frac{M_j^2}{m_\higgs^2-(\mommaj+\momlep_1)^2}  \nonumber  \\
&+16\pi\int  \lorentzd{\majneutrino_j}{\mommaj_1}\lorentzd{\lepton}{\momlep_1}
(2\pi)^4\delta(\mommaj-\momlep-\momlep_1+\mommaj_1)\nonumber\\
&\times(\momlep \momlep_1)\,\bigl[\f{\majneutrino_j}{\mommaj_1}-\f{\lepton}{\momlep_1}\bigr]
\frac{M_j^2}{m_\higgs^2-(\mommaj-\momlep_1)^2}  \kend
\end{align}
whereas contribution of the third cut is given by:
\begin{align}
\label{cutNphi}
\momlep L^{\majneutrino_j\higgs}_V & (\mommaj,\momlep) =16\pi\int 
\lorentzd{\majneutrino_j \higgs }{\mommaj_1 \momhig_1}
(2\pi)^4\delta(\mommaj_1-\momlep-\momhig_1)\nonumber \\ 
&\times (\momlep\mommaj+\momlep\momhig_1)\bigl[\f{\higgs}{\momhig_1}+\f{\majneutrino_j}{\mommaj_1}\bigr]
\frac{M_j^2}{m_\lepton^2-(\mommaj+\momhig_1)^2} \kend
\end{align}
where we have assumed $M_i<M_j$ so that the (inverse) decay $\majneutrino_i\leftrightarrow \majneutrino_j 
\lepton \lepton$ is kinematically forbidden. In \eqref{cutNl} the second term vanishes for 
the decay process $N_i \leftrightarrow \lepton \higgs$ but gives a non-zero contribution for 
the scattering processes, see section \ref{HiggsMediatedScattering}. 
If the intermediate Majorana neutrino is much heavier than the decaying one the last two cuts 
are strongly Boltzmann-suppressed. Furthermore, comparing \eqref{eqn:Lrho} and \eqref{cutlphi} we observe that 
in this case $\momlep L_V \approx \momlep  L_S$. In the same approximation we 
can also neglect the `regulator' term in the denominator of \eqref{DiagonalHermitian}. 
The two contributions to the \CP-violating parameter then have the same structure and their 
sum can be written in the form: 
\begin{align}
\epsilon_i=\epsilon_i^{vac}\,\frac{\momlep L_S}{\mommaj \momlep}\dend
\end{align}
In the vacuum limit $L^\mu_S=\mommaj^\mu$ and we recover \eqref{epsilon vacuum}.
At finite temperatures the \CP-violating parameter is moderately enhanced by the 
medium effects \cite{Frossard:2012pc}.

\section{\label{HiggsMediatedScattering} Higgs mediated scattering}

In the previous section we have approximated the full resummed Higgs propagator by 
leading-order expressions \eqref{KBHiggs} and \eqref{QPforHiggs}. In this section we 
will use a more accurate eQP approximation. As we will see, it allows one to describe 
Higgs-mediated $\Delta L=1$ two-body scattering and three-body decay processes. 

\begin{figure}[h!]
	\includegraphics[width=0.95\columnwidth]{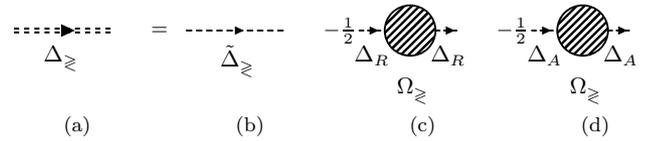}
	\caption{\label{fig:eQPHiggs}Schematic representation of the eQP approximation for the Higgs field.}
\end{figure}

Similarly to \eqref{plepeqp}, the extended quasiparticle approximation for the Higgs 
propagator reads:
\begin{equation}
\label{eqphiggs}
\pHiggs{}{\gtrless}=\pHiggsEQP{}{\gtrless}-{\textstyle\frac{1}{2}}
\left( \pHiggs{2}{R}+\pHiggs{2}{A}\right)\sHiggs{}{\gtrless}\dend
\end{equation}
Its graphic interpretation is presented in \fig\ref{fig:eQPHiggs}. 
For the first term on 
the right-hand side of \eqref{eqphiggs} we can again use approximations \eqref{KBHiggs} and 
\eqref{QPforHiggs}. To analyze the second term we have to specify the Higgs self-energy. 
At one-loop level it reads:
\begin{align}
\label{higgsse}
\sHiggs{}{\gtrless}(t,\momhig)=&g_s \yuqSqu \int 
\lorentzdd{\momQ \momtop} (2\pi)^4 \delta(\momhig-\momtop+\momQ) \nonumber\\ 
& \times \tr\bigl[ \pQ{}{\lessgtr}(t,\momQ)P_R \pTopq{}{\gtrless} (t,\momtop) P_L\bigr] \kend
\end{align}
see \app\ref{HiggsSE} for more details. As is evident from \eqref{higgsse}, here we limit 
our analysis to contributions generated by the quarks of the third generations. Let us note that in the SM the gauge 
contribution to the Higgs self-energy is of the same order of magnitude and should not be 
dismissed in a fully consistent approximation. Using the KB ansatz for the eQP 
propagators of the quarks with effective thermal mass:
\begin{subequations}
\label{eqpquark}
\begin{align}
\pTopqEQP{}{>}&=(1-f_\topq)\pTopqEQP{}{\rho} \kend \quad \pTopqEQP{}{<}=-f_\topq \pTopqEQP{}{\rho}\kend \\
\pQEQP{}{>}&=(1-f_\Q)\pQEQP{}{\rho} \kend \quad \pQEQP{}{<}=-f_\Q \pQEQP{}{\rho}\kend
\end{align}
\end{subequations}
with
\begin{subequations}
\label{spectralEQPquarks}
\begin{align}
\pQEQP{}{\rho}&=(2\pi)\sign(\momQ^0) \delta(\momQ^2-m_\Q^2 )P_L\momQslash P_R\nonumber\\
&\equiv \pQsc{}{\rho} P_L\momQslash P_R \kend  \\
\pTopqEQP{}{\rho}&=(2\pi)\sign(\momtop^0) \delta(\momtop^2-m_\topq^2 )P_R\momtopslash P_L\nonumber\\
&\equiv \pTopqsc{}{\rho} P_R\momtopslash P_L\kend 
\end{align}
\end{subequations}
and neglecting their off-shell parts, which are lepton number conserving, we can write the 
Higgs self-energy in the form:
\begin{subequations}
\label{higgssesimple}
\begin{align}
\sHiggs{}{>}(t,\momhig)&=-2g_s \yuqSqu \int \! \! 
\lorentzdd{\momQ \momtop} (2\pi)^4 \delta(\momhig+\momQ-\momtop) 
\nonumber \\ & \times \f{\Q}{} \qstatff{\topq}{} (\momQ\momtop) 
\pQsc{}{\rho}(\momQ) \pTopqsc{}{\rho}(\momtop) \kend  \\
\sHiggs{}{<}(t,\momhig)&=-2g_s \yuqSqu \int \! 
\! \lorentzdd{\momQ \momtop} (2\pi)^4 \delta(\momhig+\momQ-\momtop) 
\nonumber \\ & \times  \qstatff{\Q}{}\f{\topq}{} (\momQ\momtop) 
\pQsc{}{\rho}(\momQ) \pTopqsc{}{\rho}(\momtop) \kend
\end{align}
\end{subequations}
Substituting the one-loop lepton self-energy \eqref{Sigma1WT} with the Higgs propagator 
given by \eqref{eqphiggs} into the divergence of the lepton current \eqref{MasterEquationWigner}, 
we obtain:
\begin{align}
\label{LeptCurrSelfEn}
\frac{s{\cal H}}{z}\frac{d Y_L}{dz} & = \sum \int  \lorentzdd{\mommaj\momQ\momlep\momtop}   
(2\pi)^4 \delta(\mommaj+\momQ-\momlep-\momtop) \nonumber\\
&\times \pMajEQPsc{ii}{\rho}(\mommaj) \pLeptsc{}{\rho}(\momlep) \pQsc{}{\rho}(\momQ) 
\pTopqsc{}{\rho}(\momtop)\nonumber\\
&\times \EffAmplitude{}{\majneutrino_i\rightarrow \lepton\higgs}(\mommaj,\momlep) \pHiggs{2}{R+A}(\momtop-\momQ)
\EffAmplitude{}{\higgs\Q\rightarrow\topq}(\momtop,\momQ) \nonumber \\
&\times\bigl[\f{N_i}{\mommaj} \f{\Q}{\momQ} (1-\f{\lepton}{\momlep}) (1-\f{\topq}{\momtop}) \nonumber\\
&\hspace{18mm}-\f{\lepton}{\momlep} \f{\topq}{\momtop}(1-\f{N_i}{\mommaj})(1-\f{\Q}{\momQ})\bigr] \kend 
\end{align}
where we have introduced a combination of the retarded and advanced propagators,
\begin{align}
\label{DeltaRA}
\pHiggs{2}{R+A}(\momhig)\equiv 
{\textstyle\frac12}\bigl[\pHiggs{2}{R}(\momhig)+\pHiggs{2}{A}(\momhig) \bigr] \kend
\end{align}
which describes the intermediate Higgs line in \fig\ref{fig:NQlt} and \fig\ref{fig:NlQt}. 
Note that in \eqref{LeptCurrSelfEn} the momenta are not restricted to the mass shell. 
In particular, the zeroth components of the momenta can have either sign. Due to the 
Dirac-deltas in the spectral functions the frequency integration is trivial. Each of 
the spectral functions can be decomposed into a sum of two delta-functions, one with 
positive and one with negative frequency, leading to $2^4$ terms. These different terms 
correspond to $1 \leftrightarrow 3$ 
(inverse) decay, $2 \leftrightarrow 2$ scattering and to (unphysical) $0\leftrightarrow 4$ 
process. An additional constraint comes from the delta-function ensuring energy conservation. 
In the regime $M_i>m_\lepton + m_\Q+ m_\topq$ only $8$ terms satisfy the energy conservation. 
Using the relation 
\begin{align}
 \label{dispneg}
1 \pm \f{a}{}(t,-p)= \mp \f{\bar{a}}{} (t,p) \kend
\end{align}
where $f_{\bar{a}}$ denotes the distribution function of the antiparticles, we can then 
recast \eqref{LeptCurrSelfEn} in the form:
\begin{align}
\label{lepcurSE}
\frac{s{\cal H}}{z}\frac{d Y_L}{dz}  &= \ldots
\sum_i \int \! \! \dpi{N_i \lepton}{ \Q \topq }{\mommaj \momlep}{ \momQ \momtop} 
\\  
 \times \Bigl(&\left[\F{N_i \Q}{\lepton \topq}{\mommaj \momQ}{ \momlep \momtop} 
\EffAmpl{}{N_i \Q}{\lepton \topq}
-\F{N_i \bar \Q}{\bar \lepton \bar \topq}{\mommaj \momQ}{ \momlep \momtop} 
\EffAmpl{}{N_i \bar\Q}{\lepton \bar\topq}
\right] \nonumber \\ 
+& \left[\F{N_i \bar \topq}{\lepton \bar \Q}{\mommaj \momtop}{ \momlep \momQ}
\EffAmpl{}{N_i \bar \topq}{\lepton \bar \Q}
-\F{N_i \topq}{\bar \lepton  \Q}{\mommaj \momtop}{ \momlep \momQ}
\EffAmpl{}{N_i \topq}{\bar \lepton  \Q}
\right] \nonumber \\
+& \left[
\F{N_i \bar \lepton}{\bar \Q \topq}{\mommaj \momlep}{ \momQ \momtop} 
\EffAmpl{}{N_i \bar \lepton}{\bar \Q \topq}
-\F{N_i \lepton}{ \Q \bar\topq}{\mommaj \momlep}{ \momQ \momtop} 
\EffAmpl{}{N_i \lepton}{ \Q \bar\topq} \right] \nonumber \\ 
+&\left[
\F{N_i}{\lepton \bar \Q \topq}{\mommaj}{\momlep \momQ \momtop} 
\EffAmpl{}{N_i}{\lepton \bar \Q \topq}
-\F{N_i}{\bar \lepton \Q \bar \topq}{\mommaj}{\momlep \momQ \momtop}
\EffAmpl{}{N_i}{\bar \lepton \Q \bar \topq}
\right]
\Bigr) \nonumber\dend
\end{align}
The effective scattering amplitudes in \eqref{lepcurSE} correspond to different assignments 
for the sign of the four-momenta in \eqref{LeptCurrSelfEn}, reflecting the usual crossing symmetry. 
For the tree-level and self-energy contributions to the effective scattering and decay amplitudes 
we obtain:
\begin{subequations}
\label{EffAmplfact}
\begin{align}
\EffAmpl{T+S}{N_i \Q}{\lepton \topq}&= \EffAmpl{T+S}{N_i \bar \topq}{\lepton \bar \Q}\nonumber\\
&=\EffAmpl{T+S}{N_i}{\lepton \higgs}\Delta^2_{R+A}(\momtop-\momQ)\EffAmpl{}{\higgs\Q}{\topq}   \kend \\
\EffAmpl{T+S}{N_i \bar \lepton}{\bar \Q \topq}&=
\EffAmpl{T+S}{N_i\bar \lepton}{ \higgs}\Delta^2_{R+A}(\momtop+\momQ)\EffAmpl{}{\higgs}{\bar\Q\topq}\kend\\
\EffAmpl{T+S}{N_i }{\lepton \bar  \Q \topq} & =
\EffAmpl{T+S}{N_i}{\lepton \higgs}\Delta^2_{R+A}(\momtop+\momQ)\EffAmpl{}{\higgs}{\bar\Q\topq}\kend 
\end{align}
\end{subequations}
and similar expressions for the \CP-conjugate ones. Note that $\EffAmpl{T+S}{N_i}{\lepton \higgs}$ and 
$\EffAmpl{T+S}{N_i\bar \lepton}{ \higgs}$ are given 
by the same expression since the \CP-violating loop term in \eqref{MajoranaSelfEnAmplitudes} 
depends only on the momentum of the Majorana neutrino. In vacuum these scattering amplitudes 
reduce to \eqref{NQLtAmplitude} and \eqref{NLQtAmplitude} respectively 
but with the Feynman propagator $\pHiggs{2}{T}$ replaced by $\pHiggs{2}{R+A}$. In the latter 
the contribution of the real intermediate state is subtracted by construction \cite{Frossard:2012pc}. 
However, in the regime $m_\higgs<m_\Q+m_\topq$ the intermediate Higgs cannot be on-shell such that 
the vacuum and in-medium amplitudes become numerically equal. 
Since the amplitudes $\EffAmpl{}{\higgs}{\bar\Q\topq}$ and $\EffAmpl{}{\higgs\Q}{\topq}$ factorize 
in \eqref{EffAmplfact} and are \CP-conserving, the self-energy \CP-violating parameter in these 
processes is the same as in the Majorana decay, see \eqref{CPparamSEdecay}. However, 
since the decay and scattering processes have  different kinematics the averaged decay and 
scattering \CP-violating parameters are not  identical.

Next we consider the two-loop lepton self-energy \eqref{SEDMT}. Proceeding as above we 
find for the divergence of the lepton current an expression of the form \eqref{lepcurSE} with 
the amplitudes given by:
\begin{subequations}
\label{EffAmplvertex}
\begin{align}
\EffAmpl{V}{\majneutrino_i \Q}{\lepton \topq}&= \EffAmpl{V}{\majneutrino_i \bar \topq}{\lepton \bar \Q}\nonumber\\
&= \EffAmpl{V}{\majneutrino_i}{\lepton\higgs}\Delta^2_{R+A}(\momtop-\momQ)\EffAmpl{}{\higgs\Q}{\topq}\kend  \\
\EffAmpl{V}{\majneutrino_i \bar \lepton}{\bar \Q \topq}&=\EffAmpl{V}{N_i \bar \lepton}{\higgs} \Delta^2_{R+A}(\momtop+\momQ)
\EffAmpl{}{\higgs}{\bar\Q\topq} \kend  \\
\EffAmpl{V}{\majneutrino_i}{\lepton \bar \Q \topq} &= \EffAmpl{V}{N_i}{\lepton \higgs} \Delta^2_{R+A}(\momtop+\momQ)
\EffAmpl{}{\higgs}{\bar\Q\topq} \dend 
\end{align}
\end{subequations}
Since the vertex contribution to the Majorana decay amplitude depends on the momentum of 
the Higgs, the amplitude  $\EffAmpl{V}{N_i}{\lepton \higgs}$ does not coincide with 
$\EffAmpl{V}{N_i \bar \lepton}{\higgs}$ and we can define two inequivalent vertex \CP-violating 
parameters \cite{Nardi:2007jp}. For the scattering processes 
$\majneutrino_i \Q \leftrightarrow \lepton \topq$ and $\majneutrino_i \bar\topq  \leftrightarrow \lepton \bar \Q$ 
as well as for the three-body decay $\majneutrino_i \leftrightarrow \lepton \bar \Q \topq$,
the \CP-violating parameter coincides with \eqref{MajoranaCPVertex} with the contributions
of the three possible cuts given by \eqref{cutlphi}, \eqref{cutNl} and \eqref{cutNphi} 
respectively. For the $\majneutrino_i \bar \lepton  \leftrightarrow \bar \Q \topq$ process, 
the \CP-violating parameter still has the form \eqref{MajoranaCPVertex}, but since the 
lepton is in the initial state the loop integral must be evaluated at $(\mommaj,-\momlep)$
instead of  $(\mommaj,\momlep)$. For the first cut we obtain: 
\begin{align}
\label{cutlphi1}
\momlep L^{\lepton \higgs}_V& (\mommaj,\momlep) = 16\pi \int 
\lorentzd{\higgs \lepton}{\momhig_1 \momlep_1}
(2\pi)^4\delta(\mommaj-\momlep_1-\momhig_1) \nonumber\\ 
&\times (\momlep \momlep_1) \bigl[1+\f{\higgs}{\momhig_1}-\f{\lepton}{\momlep_1}\bigr]
\frac{M_j^2}{M_j^2-(\mommaj-\momlep_1+\momlep)^2} \dend 
\end{align}
Contributions of the second and third cuts are given by:
\begin{align}
\label{cutNl1}
\momlep L^{\majneutrino_j\lepton}_V (\mommaj,\momlep)& =16\pi\int 
\lorentzd{\majneutrino_j \lepton}{\mommaj_1 \momlep_1}
(2\pi)^4\delta(\mommaj+\momlep-\momlep_1-\mommaj_1) \nonumber\\ 
& \times(\momlep \momlep_1)\,\bigl[1-\f{\majneutrino_j}{\mommaj_1}-\f{\lepton}{\momlep_1}\bigr]
\frac{M_j^2}{(\mommaj-\momlep_1)^2-m_\higgs^2}  \nonumber  \\
&-16\pi\int  \lorentzd{\majneutrino_j \lepton }{\mommaj_1\momlep_1}
(2\pi)^4\delta(\mommaj+\momlep+\momlep_1-\mommaj_1)\nonumber\\
&\times(\momlep \momlep_1)\,\bigl[\f{\majneutrino_j}{\mommaj_1}-\f{\lepton}{\momlep_1}\bigr]
\frac{M_j^2}{(\mommaj+\momlep_1)^2-m_\higgs^2}  \kend
\end{align}
and by 
\begin{align}
\label{cutNphi1}
\momlep L^{\majneutrino_j\higgs}_V & (\mommaj,\momlep) =16\pi\int 
\lorentzd{\majneutrino_j \higgs}{\mommaj_1 \momhig_1}
(2\pi)^4\delta(\mommaj_1-\momlep-\momhig_1)\nonumber \\ 
&\times (\momlep\mommaj-\momlep\momhig_1)\bigl[\f{\higgs}{\momhig_1}+\f{\majneutrino_j}{\mommaj_1}\bigr]
\frac{M_j^2}{(\mommaj-\momhig_1)^2-m_\lepton^2} \kend
\end{align}
respectively. As follows from \eqref{cutlphi1} and \eqref{cutNl1}, \CP-violating parameter
in the $\majneutrino_i \bar \lepton  \leftrightarrow  \bar \Q \topq$ scattering receives two 
vacuum contributions \cite{Nardi:2007jp}. One is the usual cut through $\lepton \higgs$, 
and the second one is given by the first term in the cut through $N_j \lepton$. The kinematics of the 
second contribution corresponds to $N_i \lepton \leftrightarrow N_j \lepton$ $t$-channel
scattering and therefore requires the initial center-of-mass energy $s=\mommaj+\momlep$ to 
be greater than the final masses $M_j+m_\lepton$, meaning that contribution of this   
term to the reaction density is suppressed for a hierarchical mass spectrum.

\section{\label{RateEquations} Rate equations}
Solving a system of Boltzmann equations in general requires the use of numerical codes capable 
of treating large systems of stiff differential equations for the different momentum modes. This 
is a difficult task if one wants to study a wide range of model parameters. A commonly employed 
simplification is to approximate the Boltzmann equations by the corresponding system of `rate 
equations' for the abundances $Y_a$. In \cite{HahnWoernle:2009qn} it was shown that the two 
approaches, Boltzmann or the rate equations, give approximately equal results for the final 
asymmetry, up to $\sim$10\% correction. 

Starting from a quantum Boltzmann equation of the type \eqref{lepcurSE} we derive here the rate 
equation for the lepton asymmetry which includes the (usually neglected) quantum statistical 
factors. In our analysis we are closely following \cite{Frossard:2012pc}. Contribution of 
various processes to the generation of the lepton asymmetry can be represented in the form:
\begin{align}
 \label{genBoltz}
\mathcal{D}_\mu j^\mu & =  \sum_{i, \{a\},\{j\}} \int   \dpi{N_i a b \dotso}{ j k }{\mommaj 
p_a p_b \dotso}{ p_j p_k \dotso} \nonumber\\ 
&\times \bigl[ \F{N_i a b \dotso}{j k \dotso}{q p_a p_b 
\dotso}{p_j p_k \dotso} \EffAmpl{}{N_i a b \dotso}{i j \dotso}\nonumber\\
&\hspace{15mm} -\F{N_i \bar a \bar b \dotso}{\bar j \bar k \dotso}{q p_a p_b \dotso}{p_j p_k \dotso} \EffAmpl{}{N_i \bar a 
\bar b \dotso}{\bar i \bar j \dotso}\bigr] \kend 
\end{align}
compare with \eqref{lepcurSE}, where the sum runs over each possible particle state
with $\lepton \in \{j\}$ or $\bar{\lepton} \in \{a\}$. We assume 
that the SM particles are maintained in kinetic equilibrium by the fast gauge interactions. This 
means that their distribution function takes the form:
\begin{align}
 f_a(t,E_a)=\bigl(e^{\frac{E_a-\mu_a}{T}} \mp 1\bigr)^{-1}\kend
\end{align}
with a time- (or temperature-) dependent chemical potential $\mu_a=\mu_a(t)$. Here the 
upper (lower) sign corresponds to bosons (fermions). It is also useful to define the equilibrium
distribution function,
\begin{equation}
 f_a^{eq}=\bigl(e^{E_a/T}\mp 1\bigr)^{-1} \dend
\end{equation}
The fast SM interactions relate chemical potentials of the leptons, quarks and the Higgs, such 
that only one of them is independent. We can therefore express the chemical potential of the quarks 
as a function of the lepton chemical potential \cite{Buchmuller:2005eh,Kartavtsev:2005rs,PhysRevD.42.3344},
\begin{equation}
 \mu_\topq=\frac{5}{21}\mu_\lepton\equiv c_{\topq \lepton} \mu_\lepton \kend \quad \mu_\Q=
 -\frac{1}{3}\mu_\lepton\equiv c_{\Q \lepton} \mu_\lepton \dend
\end{equation}
Chemical potentials of the antiparticles:  $\mu_{\bar a}=-\mu_a$. The lepton chemical potential 
is related to the abundance by:
\begin{align}
 \frac{\mu_\lepton}{T} \approx c_\lepton \frac{Y_L}{2Y_\lepton^{eq}} \kend
\end{align}
where $c_\lepton$ depends on the thermal mass of the lepton. For $m_\lepton/T \approx 0.2$ it can 
be very well approximated by the zero mass limit, $c_\lepton \approx 9 \xi(3)/\pi^2 \approx 1.1$. 

Using the identity $1 \pm f_a=e^{(E_a-\mu_a)/T} f_a$ and energy conservation we can rewrite 
the combinations of distribution functions appearing in \eqref{genBoltz} as:
\begin{align}
 \label{Fsimplif}
&\F{N_i a b \dotso}{j k \dotso}{q p_a p_b \dotso}{p_j p_k \dotso}= \\ 
& \times (2\pi)^4\delta\left(q+{\textstyle\sum}_a p_a-{\textstyle\sum}_j p_j\right)  
\frac{\prod_a f_a \prod_j(1\pm f_j)}{1-f_{N_i}^{eq}} \nonumber \\
&\times \bigl[ f_{N_i}-f_{N_i}^{eq}-f_{N_i}^{eq}(1-f_{N_i})\{ e^{\sum_j\mu_j/T-\sum_a \mu_a/T}-1\}\bigr]\kend \nonumber
\end{align}
where we have suppressed the momentum argument in the distribution functions for notational 
convenience. We can then expand \eqref{Fsimplif} in the small chemical potential $\mu_a$. 
Assuming the Majorana neutrino to be close to equilibrium, $f_{N_i}-f_{N_i}^{eq}\sim \mathcal{O}(\mu_a)$, 
we see that the term in the square bracket in \eqref{Fsimplif} is already of the first order in the chemical 
potential. We can therefore replace the distribution functions in the second line of \eqref{Fsimplif} 
by the equilibrium ones,
\begin{align}
 \frac{\prod_a f_a^{eq} \prod_j(1\pm f_j^{eq})}{1-f_{N_i}^{eq}}=&\prod_a f_a^{eq} \prod_j(1 \pm f_j^{eq})\nonumber \\ 
 &+\prod_j f_j^{eq} \prod_a (1\pm f_a^{eq} ) \kend
\end{align}
and expand the exponential to first order in the chemical potential. Since we assume small 
departure from equilibrium the Majorana distribution function that multiplies the chemical 
potential should also be replaced by the equilibrium one. The corresponding equation for 
the antiparticles is obtained from the above equation by replacing $\mu_a\rightarrow -\mu_a$. 
The last step is to assume that the Majorana distribution function is proportional to its equilibrium value,
\begin{align}
 f_{N_i} \approx \frac{Y_{N_i}(t)}{Y_{N_i}^{eq}(t)}f_{N_i}^{eq} \dend
\end{align}
Putting everything together we get the conventional form of the rate equation,
\begin{align}
\label{rateeq}
\frac{s{\cal H}}{z}\frac{d Y_L}{dz}&=
\sum_{i, \{a\},\{j\}} 
\left[\langle \epsilon^{N_i a b \dotso}_{j k \dotso}
\gamma^{N_i a b \dotso}_{ j k \dotso}\rangle 
\left(\frac{Y_{N_i}}{Y_{N_i}^{eq}}-1\right) \right.\nonumber \\ 
&-\left.\langle \gamma^{N_i a b \dotso}_{j k \dotso} \rangle 
c_\lepton c_{a b \dotso \leftrightarrow j k \dotso}\frac{Y_L}{2Y_L^{eq}}\right] \kend
\end{align}
where we have defined the production and washout reaction densities:
\begin{subequations}
 \label{LeptonReactDens}
\begin{align}
\label{reaceps}
&\langle \epsilon^{N_i a b \dotso}_{j k \dotso}
\gamma^{N_i a b \dotso}_{j k \dotso}\rangle  \equiv  \nonumber \\ 
&\equiv \int \dpi{N_i a b \dotso}{j k \dotso }{\mommaj p_a p_b \dotso}{ p_j p_k \dotso}
(2\pi)^4 \delta\left(q+{\textstyle\sum}_a p_a-{\textstyle\sum}_j p_j\right) \nonumber\\
&\times \epsilon_{N_i a b \dotso \rightarrow j k \dotso} \left(\EffAmplitude{}{N_i a b \dotso \leftrightarrow j k \dotso}+
\EffAmplitude{}{N_i \bar a \bar b \dotso \leftrightarrow  \bar j \bar k \dotso}\right) f_{N_i}^{eq} \nonumber \\ 
&\times \Bigl( \prod_a f_a^{eq} \prod_j(1 \pm f_j^{eq})+\prod_j f_j^{eq}
\prod_a (1\pm f_a^{eq} ) \Bigr)\kend \\
\label{WashoutScattering}
&\langle \gamma^{N_i a b \dotso}_{j k \dotso} \rangle \big|_W \equiv\nonumber\\
& \equiv \int 
\dpi{N_i a b \dotso}{ j k \dotso}{\mommaj p_a p_b \dotso}{ p_j p_k \dotso} 
(2\pi)^4\delta\left(q+{\textstyle\sum}_a p_a-{\textstyle\sum}_j p_j\right)\nonumber \\ 
& \times  \left(\EffAmplitude{}{N_i a b \dotso \leftrightarrow  j k \dotso}+
\EffAmplitude{}{N_i \bar a \bar b \dotso \leftrightarrow   \bar j \bar k \dotso}\right) 
f_{N_i}^{eq}(1-f_{N_i}^{eq}) \nonumber \\ & 
\times  \Bigl( \prod_a f_a^{eq} 
\prod_j(1 \pm f_j^{eq})+\prod_j f_j^{eq} \prod_a (1\pm f_a^{eq} ) \Bigr)\kend
\end{align}
\end{subequations}
and the numerical factor,
\begin{equation}
 \label{defcc}
c_{a b \dotso \leftrightarrow j k \dotso}\equiv\frac{\sum_j \mu_j-\sum_a \mu_a}{\mu_\lepton}\dend
\end{equation}
Equation \eqref{rateeq} must be supplemented by an equation for the heavy neutrino abundance,
\begin{align}
 \label{majneutab}
\frac{s{\cal H}}{z}&\frac{d Y_{N_i}}{dz}= -\sum_{\{a\},\{j\}} \langle \gamma^{N_i a b \dotso}_{j k \dotso} 
\rangle \big|_P \left(\frac{Y_{N_i}}{Y_{N_i}^{eq}}-1\right) \kend
\end{align}
with the reaction density given by an expression similar to \eqref{WashoutScattering}:
\begin{align}
\label{MajReactDens}
&\langle \gamma^{N_i a b \dotso}_{j k \dotso} \rangle \big|_P \equiv\nonumber\\
& \equiv \int 
\dpi{N_i a b \dotso}{ j k \dotso}{\mommaj p_a p_b \dotso}{ p_j p_k \dotso} 
(2\pi)^4\delta\left(q+{\textstyle\sum}_a p_a-{\textstyle\sum}_j p_j\right)\nonumber \\ 
& \times  \left(\EffAmplitude{}{N_i a b \dotso \leftrightarrow  j k \dotso}+
\EffAmplitude{}{N_i \bar a \bar b \dotso  \leftrightarrow   \bar j \bar k \dotso}\right) 
f_{N_i}^{eq} \nonumber \\ & 
\times  \Bigl( \prod_a f_a^{eq} 
\prod_j(1 \pm f_j^{eq})+\prod_j f_j^{eq} \prod_a (1\pm f_a^{eq} ) \Bigr)\dend
\end{align}
Note that these expressions are valid for two-body scattering processes with Majorana
neutrino in the initial or final state as well as for Majorana (inverse) decay processes.

If the quantum-statistical corrections are neglected, i.e. if the $1\pm \f{}{}$ terms are 
replaced by unity and the equilibrium fermionic and bosonic distributions are approximated 
by the Maxwell-Boltzmann one, then \eqref{WashoutScattering} and \eqref{MajReactDens} are 
equal. For the case of a $2\leftrightarrow 2$ scattering process they read:
\begin{align}
\label{WashoutScatteringMB}
\langle \gamma^{N_i a}_{j k} \rangle &  \equiv
\int 
\dpi{N_i a}{ j k }{\mommaj p_a}{ p_j p_k} (2\pi)^4\delta (q+p_a-p_j-p_k)\nonumber \\ 
& \times  \bigl(\EffAmplitude{}{N_i a  \leftrightarrow j k}+
\EffAmplitude{}{N_i \bar a  \leftrightarrow \bar j \bar k}\bigr)
f_{N_i}^{eq} f_a^{eq}\dend
\end{align}
Part of the integrations in \eqref{WashoutScatteringMB} can be performed analytically and we obtain:
\begin{align}
	\label{ReactDensCanon}
	\langle \gamma^{N_i a}_{j k}\rangle & \approx \frac{T}{64\pi^4}\int\limits_{s_{min}}^\infty ds 
	\sqrt{s}K_1\left(\frac{\sqrt{s}}{T}\right) \hat{\sigma}^{N_i a }_{j k }(s)\dend
\end{align}
Here $s_{min}=(M_i+m_a)^2$ (assuming $M_i+m_a>m_j+m_k$) and $\hat \sigma(s)$ is the so-called reduced cross-section:
\begin{align}
\label{eqn:reduced cross section definition main text}
\hat \sigma^{N_i a }_{j k } \equiv \frac{1}{8\pi}\int\limits_0^{2\pi} 
	\frac{d\varphi_{ai}}{2\pi}
	\int\limits_{t^-}^{t^+} \frac{dt}{s}\,
	\bigl(\EffAmplitude{}{N_i a  \leftrightarrow  j k }{} +
	\EffAmplitude{}{N_i \bar a  \leftrightarrow  \bar j \bar k }{}\bigr)
	\kend
\end{align}
where $s$ and $t$ are the usual Mandelstam variables. The integration limits are given by
\cite{Frossard:2012pc}:
\begin{align}
\label{eqn: t range}
t^\pm&=M_i^2+m_j^2\nonumber\\
&-\frac{s}{2}\bigl[(1+M_i^2/s-m_a^2/s)(1+m_j^2/s-m^2_k/s)\nonumber\\
&\mp \lambda^\frac12(1,M_i^2/s,m_a^2/s)\lambda^\frac12(1,m_j^2/s,m_k^2/s)\bigr]\kend
\end{align}	
where $\lambda(x,y,z)\equiv x^2+y^2+z^2-2xy-2xz-2yz$ is the usual kinematical function. 
If effective thermal masses of the SM particles are neglected then the integration limits 
simplify to $t^+=0$ and $t^-=-(s-M_i^2)$. Integrating \eqref{NQLtAmplitude} and 
\eqref{NLQtAmplitude} over $t$ and neglecting the effective masses of the initial and final 
lepton and quarks we obtain the standard expressions (see, e.g. \cite{Luty:1992un,Plumacher:1998ex}) 
for the reduced cross-sections of the Higgs-mediated scattering processes:
\begin{subequations}
\label{VacuumReducedCrs}
\begin{align}
\label{NQLtCanRedAmpl}
\hat{\sigma}^{N_i \Q}_{\lepton\topq}&=\sigma^{N_i \bar\topq}_{\lepton\bar\Q}=\frac{g_w g_s}{4\pi} 
(\yu^\dagger \yu)_{ii}\, \yuqSqu\,\frac{x-a_i}{x}\\
&\times\left[\frac{x-2a_i+2a_\higgs}{x-a_i+a_\higgs}
+\frac{a_i-2a_\higgs}{x-a_i}\ln\left(\frac{x-a_i+a_\higgs}{a_\higgs}\right)\right]\kend\nonumber\\
\label{NLQtCanRedAmpl}
\hat{\sigma}^{N_i \bar\lepton}_{\bar\Q\topq}&=\frac{g_w g_s}{4\pi} 
(\yu^\dagger \yu)_{ii}\yuqSqu \frac{(x-a_i)^2}{(x-a_\higgs)^2}\kend
\end{align}
\end{subequations}
where we have replaced $s$ by $x\equiv s/M_1^2$ and introduced dimensionless quantities 
$a_i\equiv M_i^2/M_1^2$ and $a_\higgs\equiv m_\higgs^2/M_1^2$. Combined with \eqref{ReactDensCanon},
expressions \eqref{VacuumReducedCrs} give the conventional reaction densities of the 
Higgs-mediated scattering processes.

Since in the conventional approach the \CP-violating parameter is calculated in vacuum 
it is momentum-independent and therefore can be taken out of the integral. The \CP-violating 
reaction densities are thus proportional to the washout ones: 
\begin{align}
\langle \epsilon^{N_i a}_{jk} \gamma^{N_i a}_{jk}\rangle&=\epsilon_i^{vac}\langle \gamma^{N_i a}_{jk}\rangle\kend 
\end{align}
where we have again assumed a strongly hierarchical mass spectrum of the heavy neutrinos. 
When the medium corrections are taken into account the \CP-violating parameter depends on the 
momenta of the initial and final states and this simple relation is violated. 

\section{\label{Numerics}Numerical results}

To illustrate the effect of the quantum-statistical corrections and effective thermal 
masses  we present in this section ratios of the reaction densities to the conventional ones
assuming  a strongly hierarchical mass spectrum of the Majorana neutrinos. 

\begin{figure}[t]
\includegraphics[width=0.95\columnwidth]{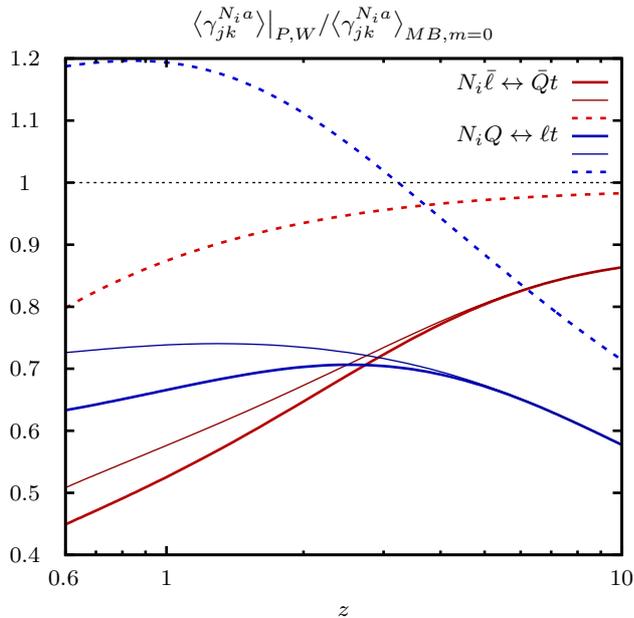}
\caption{\label{RatioWashoutReactDens} Ratios of the scattering reaction densities 
	obtained taking into account the thermal masses (dashed lines) and the thermal
	masses plus quantum-statistical effects (solid lines) to the conventional ones.
	The thick solid lines correspond to \eqref{WashoutScattering} whereas the thin 
	ones to \eqref{MajReactDens}.}
\end{figure}

\begin{figure}[t]
\includegraphics[width=0.95\columnwidth]{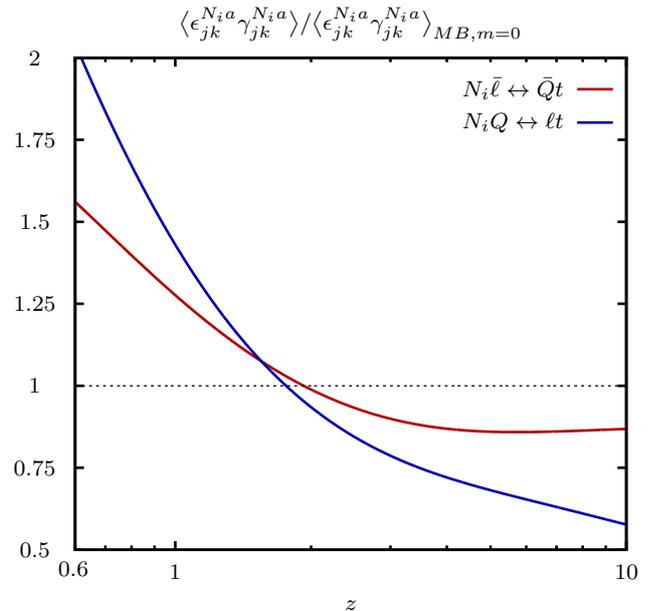}
\caption{\label{RatioCPReactDens} Ratio of the \CP-violating reaction densities to the 
ones computed using Boltzmann statistics and neglecting the thermal masses of the initial 
and final states.}
\end{figure}

Let us first consider the scattering processes.
Ratios of the improved reaction densities to the conventional ones are presented 
in \fig\ref{RatioWashoutReactDens}. 
Note that the Majorana (as well as the quark) Yukawa couplings cancel out in these ratios 
and for this reason we do not specify them here. The dashed lines show the ratio of the 
reaction density computed using  \eqref{ReactDensCanon}--\eqref{eqn: t range}, i.e. taking 
into account the effective thermal masses but neglecting the quantum-statistical corrections, 
to the conventional ones. For the $\majneutrino_i\bar\lepton\leftrightarrow \bar\Q\topq$ 
process (dashed red line) the effective 
masses decrease the available phase space and lead to a suppression of 
the reaction density in the whole range of temperatures. Note that the ratio does not 
approach unity at low temperatures. Qualitatively this behavior can be understood from 
\eqref{WashoutScatteringMB}. Let us assume for a moment that the scattering amplitude is 
momentum-independent. The reaction density at low temperatures can then be estimated by 
evaluating the distribution functions at the average momenta $\langle p_i\rangle$ and 
$\langle p_a\rangle\sim 3T$.  In the ratio of the reaction densities the Majorana 
distribution function cancels out and 
\begin{align*}
\frac{\langle \gamma^{X}_{Y}\rangle_{MB,m\neq 0}}{\langle \gamma^{X}_{Y}\rangle_{MB,m=0}}
&\approx \frac{\exp(-E_a/T)}{\exp(-\langle p_a\rangle/T)} \approx \exp(-m_a^2/2\langle p_a\rangle T)\dend
\end{align*} 
A more accurate estimate for the ratio of $\langle \gamma^{X}_{Y}\rangle_{MB,m\neq 0}$  and 
$\langle \gamma^{X}_{Y}\rangle_{MB,m=0}$ is $\sim \exp(-m_a^2/T^2)$. Since  to a good 
approximation  $m_a \propto T$ we conclude that this ratio is a constant smaller than unity.
In other words, despite the fact that at low temperatures the quark masses become small compared to the 
Majorana mass this ratio is \textit{not} expected to approach unity as the temperature decreases. Note also 
that (in a very good agreement with the numerical cross-check) this ratio does not depend on the masses 
of the final states. Of course, the momentum dependence of the scattering amplitude somewhat changes the low-temperature  
behavior of the reaction density. Interestingly enough, for the $\majneutrino_i\Q\leftrightarrow\lepton\topq$
process the inclusion of the thermal masses actually enhances the reaction density at high temperatures 
(dashed blue line). This occurs because the induced increase of the amplitude turns out to be larger 
than the phase-space suppression. At low temperatures the effective masses become negligible in the 
scattering amplitude but still play an important role in the kinematics. As a result, the ratio becomes 
smaller than unity and continues to decrease as the temperature decreases. Let us note that for a 
(moderately) strong washout regime most of the asymmetry is typically produced at $z \lesssim 10$
and the low-temperature behavior of the reaction densities does not affect the generation of the 
asymmetry. Since all particles in the initial and final states are fermions  the quantum-statistical 
effects further suppress the reaction densities (solid blue and red lines) and render the ratio of the 
improved and conventional reaction densities smaller than unity for both $\majneutrino_i\Q\leftrightarrow\lepton\topq$ 
and $\majneutrino_i\bar \lepton\leftrightarrow \bar \Q\topq$ in the whole range of temperatures. 
Ratios of the improved \CP-violating scattering reaction densities to the conventional ones are 
presented in \fig\ref{RatioCPReactDens}.
For both scattering processes the improved \CP-violating reaction densities are enhanced at high 
temperatures. This is explained by the enhancement of the \CP-violation in the Majorana decay observed 
in \cite{Garny:2010nj,Frossard:2012pc}. At the intermediate temperatures the relative enhancement of the \CP-violating 
parameters gets smaller and is overcompensated by the effective mass and Fermi-statistics induced 
suppression of the washout reaction densities that we have observed in \fig\ref{RatioWashoutReactDens}.
The low-temperature behavior is somewhat different for the two scattering processes. For the 
$N_i \bar\lepton \leftrightarrow \bar\Q \topq$ process the effective mass and quantum-statistical 
effects get smaller in both the (unintegrated) \CP-violating parameter and the washout reaction 
density, such that the ratio of the \CP-violating reaction density to the 
conventional one slowly approaches a constant value. 
On the other hand, for the $N_i \Q\leftrightarrow\lepton \topq$ process the suppression of the 
washout reaction density induced by the effective masses of the initial and final states that 
we observed in \fig\ref{RatioWashoutReactDens} also leads to a suppression of the \CP-violating 
reaction density that gets stronger at low temperatures.

\begin{figure}[t]
\includegraphics[width=0.95\columnwidth]{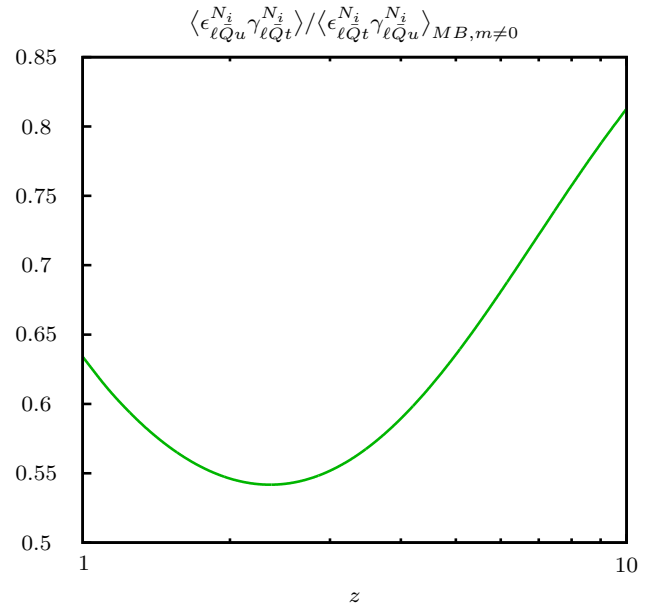}
\caption{\label{RatioCPDecayReactDens} Ratio  of the \CP-violating reaction density of the 
	$\majneutrino_i \leftrightarrow  \lepton \bar\Q\topq$ process obtained taking into account 
	effective thermal masses and quantum-statistical effects to the ones
	computed taking into account only the effective thermal masses.}
\end{figure}

Next we consider the three-body decay. 
\begin{figure}[t]
\includegraphics[width=0.95\columnwidth]{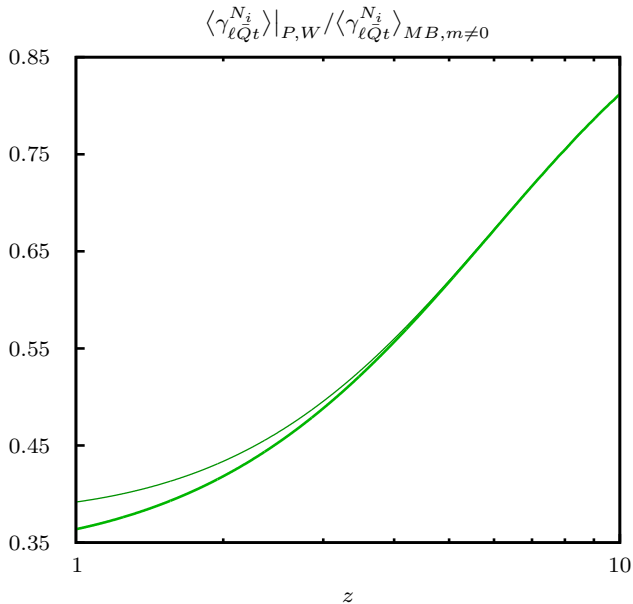}
\caption{\label{RatioDecayReactDens} Ratio  of the $\majneutrino_i \leftrightarrow  \lepton \bar\Q\topq$
decay reaction density 	obtained taking into account effective thermal masses   and quan\-tum-statistical 
effects to the conventional one  computed taking into account only the effective thermal masses of the 
final and intermediate states. The thick solid line corresponds to \eqref{LeptonReactDens} whereas 
the thin one to \eqref{MajReactDens}.}
\end{figure} 
Neglecting the quantum-statistical effects and using vacuum 
approximation for the $\majneutrino_i \leftrightarrow \lepton \bar\Q\topq$ decay amplitude in  
\eqref{LeptonReactDens} and \eqref{MajReactDens} we recover the conventional expression for the decay 
reaction density:
\begin{align}
	\label{DecayReactDensCanon}
	\langle \gamma^{N_i}_{\lepton\bar\Q\topq}\rangle & \approx 
	\frac{g_N}{2\pi^2}T M_i^2 \Gamma_{N_i\rightarrow \lepton\bar\Q\topq}
	K_1(M_i/T)\dend
\end{align}
Note that it is important to retain the effective thermal masses of the quarks in the calculation 
of $\Gamma_{N_i\rightarrow \lepton\bar\Q\topq}$. The four-momentum of the intermediate 
Higgs, see \fig\ref{fig:Ampl_N_lQt},  varies in the range $(m_\Q+m_\topq)^2\leqslant \momhig^2 \leqslant(M_i-m_\lepton)^2$. 
The relation $m_\higgs < m_\Q+m_\topq$, which is fulfilled in the SM, ensures that the intermediate 
Higgs remains off-shell and prevents a singularity in the Higgs propagator.
The ratio of the reaction density computed taking into account the quantum-statistical effects 
and effective masses to the one computed taking into account only the effective masses is 
presented in \fig\ref{RatioDecayReactDens}. 
Note that since $m_\Q\approx m_\topq \approx 0.4\, T$
and $m_\lepton \approx 0.2\, T$ this three-body decay is kinematically allowed only at $T \lesssim M_i$.
As one would expect, at high temperatures the fermionic nature of the initial and final states 
leads to a suppression as compared to the Boltzmann approximation. At low temperatures the 
quantum-statistical effects play no role and the ratio slowly approaches unity.
Ratio of the \CP-violating reaction density for the $N_i \leftrightarrow  \lepton\bar\Q\topq$ process 
is presented in \fig\ref{RatioCPDecayReactDens}. 
At high and intermediate temperatures the medium 
enhancement of the (unintegrated) \CP-violating parameter is overcompensated by the suppression of 
the washout decay reaction density that we have observed in \fig\ref{RatioDecayReactDens}.
At low temperatures the effective mass and quantum-statistical effects get smaller in both 
the (unintegrated) \CP-violating parameter and the washout reaction density, such that the 
\CP-violating reaction density slowly approaches the conventional one. 

\begin{figure}[ht]
\includegraphics[width=0.95\columnwidth]{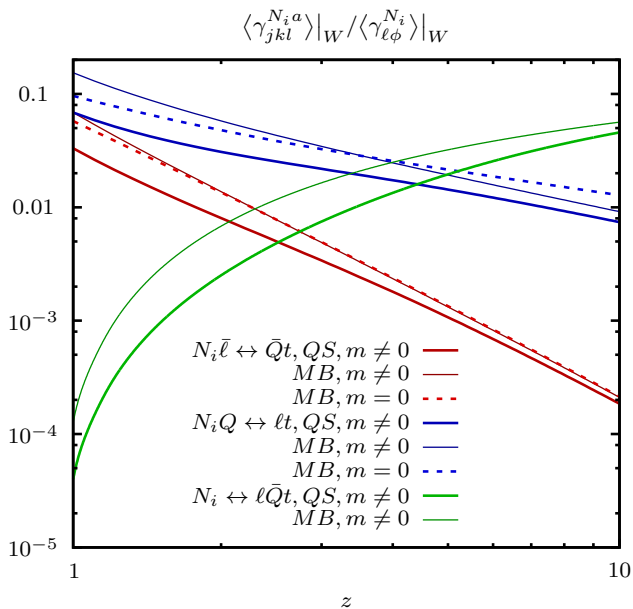}
\caption{\label{ScatToDecay} Ratio of the washout scattering and three-body decay reaction densities
to the reaction density of $\majneutrino_i \leftrightarrow  \lepton\higgs$ process. The dashed lines 
denote the ratios of the conventional reaction densities, the thin solid lines the ratios 
computed taking into account only the effective masses in all the reaction densities, and finally the thick solid lines the 
ratios computed taking into account the effective masses and quantum-statistical corrections 
in all the reaction densities. The reaction density $\langle \gamma^{N_i}_{\ell\phi} \bigr\rangle_{W}$
is computed using \eqref{EffAmplMajDecay} and the definition \eqref{WashoutScattering}, see also \cite{Frossard:2012pc}.}
\end{figure}

To conclude this section we present the ratio of the three-body decay and $2\leftrightarrow 2$
scattering processes to the reaction density of $\majneutrino_i \leftrightarrow  \lepton\higgs$ 
process, see \fig\ref{ScatToDecay}.
As can be inferred from this plot, the three-body decay is subdominant in the whole range of 
temperatures and can be safely neglected. The inclusion of the effective masses has a very 
similar effect on the two-body decay and scattering reaction densities such that the ratios 
of the two almost do not change as compared to the one computed in the massless approximation. 
The inclusion of the quantum-statistical corrections has a stronger effect on the scattering
processes such that the ratio of the reaction densities is smaller than the ratio of the 
conventional ones. Let us also note that the scattering processes are very important at high
temperatures but become subdominant at low temperatures.

\section{\label{Summary}Summary}

In this work we have studied $\Delta L=1$ decay and scattering processes mediated by the Higgs 
with quarks in the initial and final states using the formalism of non-equilibrium quantum field 
theory. 

Starting from the Kadanoff-Baym equations for the lepton propagator we have derived the 
corresponding qu\-a\-ntum-corrected Boltzmann and rate equations for the total lepton asymmetry. 
As compared to the canonical ones  the latter are free of the notorious double-counting 
problem and ensure that the asymmetry automatically vanishes in thermal equilibrium. To 
compute the collision term we have taken into account one- and two-loop contributions 
to the lepton self-energy and used the extended quasiparticle approximation for the Higgs 
propagator. The impact of the SM gauge interactions on the collision term has been approximately
taken into account in the form of effective thermal masses of the Higgs, leptons and quarks.

We find that the inclusion of the effective masses and quantum-statistical terms 
suppresses the washout reaction densities of the decay and scattering processes with 
respect to the conventional ones, where these effects are neglected, in the whole 
relevant range of temperatures. For the $N_i \bar\lepton \leftrightarrow \bar\Q \topq$ process 
the ratio of the improved and conventional washout reaction densities slowly approaches a constant 
value close to unity at low temperatures. Interestingly enough, for the $N_i \Q\leftrightarrow \lepton \topq$ 
processes this ratio decreases even at low temperatures. Finally for $N_i\leftrightarrow \lepton\bar\Q \topq$ 
process the ratio slowly approaches unity at low temperatures. As far as the \CP-violating reaction densities
are concerned, we find that for the scattering processes the ratio of the improved and 
the conventional ones is greater than unity at high temperatures but is smaller than
unity at intermediate and low temperatures because of the thermal masses and 
quantum-statistical effects. For the three-body decay this ratio is smaller than 
unity in the whole relevant range of temperatures.

We expect that the effects studied here can induce a $\mathcal{O}(10 \%)$ correction to the total 
generated asymmetry. For a detailed phenomenological analysis it is necessary to include 
further phenomena such as flavour effects and process with gauge bosons in the initial 
and final states.

\subsection*{Acknowledgements}
\noindent  The work of A.K. has been supported by the German Science Foundation (DFG) 
under Grant KA-3274/1-1 ``Systematic analysis of baryogenesis in non-equilibrium quantum 
field theory''. T.F. acknowledges support by the IMPRS-PTFS. We thank A.~Hohenegger for 
useful discussions.

\begin{appendix}
\section{\label{HiggsSE}Higgs self-energy}
The top quark contribution to the Higgs self-energy is derived from the 2PI effective action. At 
one-loop the contribution of the top quark is given by:
\begin{align}
 i\Gamma_2 = g_s \yuqSqu \int_\mathcal{C} \! d^4u d^4v &
 \tr \left[ \pQ{}{ba}(v,u)P_R \pTopq{}{}(u,v) \right] \nonumber \\ 
 &\times \epsilon^*_{bc} \pHiggs{}{cd}(v,u) \epsilon_{da}^T\kend
\end{align}
where the factor $g_s=3$ comes from the summation over color indices and $\epsilon=i\sigma_2$. 
In a $SU(2)_L$ symmetric state  the Higgs and lepton propagators are proportional to the identity 
in the $SU(2)_L$ space, and so is the Higgs  self-energy,
\begin{align}
 \sHiggs{}{ab}&(x,y)\equiv \sHiggs{}{}(x,y) \delta_{ab}=\frac{\delta i\Gamma_2 }{\pHiggs{}{ba}(y,x)} \nonumber \\
 &=g_s \yuqSqu \tr \left[ \pQ{}{}(y,x)P_R \pTopq{}{}(x,y)P_L \right] \delta_{ab} \dend
\end{align}
Its Wightman components are given by,
\begin{align}
 \sHiggs{}{\gtrless}(x,y)=g_s \yuqSqu \tr 
 \bigl[ \pQ{}{\lessgtr}(y,x)P_R \pTopq{}{\gtrless}(x,y)P_L \bigr]\dend
\end{align}
Finally, performing a Wigner transform of the above equation, we find,
\begin{align}
 \sHiggs{}{\gtrless}(t,k)&=g_s\yuqSqu  \int 
 \lorentzdd{\momQ \momtop} (2\pi)^4 \delta(k-\momtop+\momQ)\nonumber \\ 
 & \times  \tr \bigl[\pQ{}{\lessgtr}(t,\momQ)P_R \pTopq{}{\gtrless}(t,\momtop)P_L \bigr]\dend 
\end{align}

\section{\label{kinematics}Reaction density of \texorpdfstring{$1\rightarrow 3$}{1->3} decay}

For $\majneutrino_i\rightarrow \lepton \bar \Q \topq$ decay the general  expression \eqref{MajReactDens}
takes the form: 
\begin{align}
\bigl\langle \gamma^{N_i}_{\lepton \bar \Q \topq}\bigr\rangle & =
\int \dpi{N_i }{\lepton \bar \Q \topq}{\mommaj}{\momlep\momQ\momtop} (2\pi)^4 \delta(\mommaj-\momlep-\momQ-\momtop)\\
&\times\EffAmplitude{}{\majneutrino_i\rightarrow \lepton\higgs}\times \pHiggs{2}{R+A}(\momQ+\momtop)\times
\EffAmplitude{}{\higgs\rightarrow \bar\Q\topq}\nonumber\\
& \times \f{\majneutrino_i}{eq}\bigl[
\qstatff{\lepton}{eq}\qstatff{\bar\Q}{eq}\qstatff{\topq}{eq}+\f{\lepton}{eq}\f{\bar\Q}{eq}\f{\topq}{eq}
\bigr]\kend\nonumber
\end{align}
where we have used the explicit form of the decay amplitude \eqref{NtoLQtAmplitude}. To reduce it to a 
form suitable for the numerical analysis we insert an identity:
\begin{align}
	1=\int ds \int d^4 k \delta(\momQ+\momtop-\momhig)\delta_+(\momhig^2-s)\kend
\end{align}
where $\momhig$ is the four-momentum of the intermediate Higgs. Approximating furthermore $\pHiggs{2}{R+A}$
by $\pHiggs{2}{T}$ we can rewrite the reaction density in the form:
\begin{align}
\label{NtoLQtReactDensPrel}
\bigl\langle \gamma^{N_i}_{\lepton \bar \Q \topq}\bigr\rangle & =
\int \dpi{N_i}{}{\mommaj}{} \,\f{\majneutrino_i}{eq}\,\int\frac{ds}{2\pi} \pHiggs{2}{T}(s)\\
&\times \int \dpi{}{\lepton \higgs}{}{\momlep\momhig} \,
(2\pi)^4 \delta(\mommaj-\momlep-\momhig)  \EffAmplitude{}{\majneutrino_i\rightarrow \lepton\higgs}\nonumber\\
&\times \int \dpi{}{\bar \Q \topq}{}{\momQ\momtop} (2\pi)^4 \delta(\momhig-\momQ-\momtop)\,
\EffAmplitude{}{\higgs\rightarrow \bar\Q\topq}\nonumber\\
& \times \bigl[
\qstatff{\lepton}{eq}\qstatff{\bar\Q}{eq}\qstatff{\topq}{eq}+\f{\lepton}{eq}\f{\bar\Q}{eq}\f{\topq}{eq}
\bigr]\dend\nonumber
\end{align}
Note that in the regime $m_\higgs<m_\Q+m_\topq$ realized in the considered 
case the Higgs is always off-shell and its width can be neglected in $\pHiggs{}{T}$. For the 
second line in \eqref{NtoLQtReactDensPrel} we can use 
\cite{Frossard:2012pc}:
\begin{align}
\label{ellphiintegral}
\int \dpi{}{\lepton \higgs}{}{\momlep\momhig}  (2\pi)^4& \delta(\mommaj-\momhig-\momlep)\nonumber\\
& \rightarrow \frac{1}{8\pi\,|\vec \mommaj|} \int\limits_{E^{-}_\momlep}^{E^{+}_\momlep}dE_\momlep
\int\limits_0^{2\pi}\frac{d\varphi}{2\pi}\dend
\end{align}
The integration limits are given by 
\begin{align}
\label{eqn: decay integration limits}
E^\pm_\momlep =
{\textstyle\frac12}\bigl[E_\mommaj\bigl(1+x_\lepton-x_\higgs)
\pm |\vec \mommaj| \lambda^\frac12(1,x_\lepton,x_\higgs)\bigr]\kend
\end{align}
where $x_\lepton\equiv m_\lepton^2/M_i^2$, $x_\higgs\equiv s/M_i$ and 
$\lambda(x,y,z)\equiv x^2+y^2+z^2-2xy-2xz-2yz$ is the usual kinematical function. 
For the third line we can use a similar expression with $x_\Q=m_\Q^2/s$ and 
$x_\topq=m_\topq^2/s$. 

Expressed in terms of the integration variables the amplitudes take the form:
\begin{subequations}
\begin{align}
\EffAmplitude{}{\majneutrino_i\rightarrow\lepton \higgs} 
& = g_w (h^\dagger h)_{ii} (M_i^2+m_\lepton^2-s)\kend\\
\EffAmplitude{}{\higgs \rightarrow \bar \Q\topq} & =  
g_s \yuqSqu (s-m_\Q^2-m_\topq^2)\dend
\end{align}
\end{subequations}
Since they do not depend on the angles between the quarks and leptons the integration over 
$\varphi$ is trivial and the reaction density takes the form:
\begin{align}
\label{NtoLQtReactDens}
\bigl\langle \gamma^{N_i}_{\lepton \bar \Q \topq}\bigr\rangle & =
\int \dpi{N_i}{}{\mommaj}{} \,\f{\majneutrino_i}{eq}\,\int\frac{ds}{2\pi} \pHiggs{2}{T}(s)\\
&\times \int^{E^+_\momlep}_{E^-_\momlep} \frac{dE_\momlep}{8\pi |\vec{\mommaj}|} \, 
\EffAmplitude{}{\majneutrino_i\rightarrow \lepton\higgs}
\int^{E^+_\momQ}_{E^-_\momQ} \frac{dE_\momQ}{8\pi |\vec{\momhig}|} \, 
\EffAmplitude{}{\higgs \rightarrow \bar \Q\topq}
\nonumber\\
& \times \bigl[
\qstatff{\lepton}{eq}\qstatff{\bar\Q}{eq}\qstatff{\topq}{eq}+\f{\lepton}{eq}\f{\bar\Q}{eq}\f{\topq}{eq}
\bigr]\dend\nonumber
\end{align}
The three-momentum of the intermediate Higgs is given by $|\vec{\momhig}|=(E_\momhig^2-s)^\frac12$
and $E_\momhig=E_\mommaj-E_\momlep$. Note that if we  neglect the quantum-statistical factors 
in \eqref{NtoLQtReactDens} the reaction density takes the standard form.

\section{Majorana spectral self-energy}

We compute here the Majorana spectral self-energy. In a \CP-symmetric medium it reads \cite{Frossard:2012pc}:
\begin{align}
 \sMaj{ij}{\rho} &=-{\frac{g_w}{16\pi}}\bigl[(h^\dagger h)_{ij}P_L+(h^\dagger h)^*_{ij} P_R\,\bigr] L_{S} \kend
\end{align}
where we have defined the loop function $L_S(q)$,
\begin{align}
\label{defLrhoh}
 L_{S}(q)=&16\pi\int  \! \! \lorentzdd{\momlep \momhig} (2\pi)^4\delta(\mommaj-\momlep-\momhig)\slashed{\momlep} \nonumber \\
 &\times \big[ \pHiggs{}{F}(\momhig) \pLeptsc{}{\rho(h)}(\momlep)+\pHiggs{}{\rho(h)}(\momhig) \pLeptsc{}{F}(\momlep)\big] \dend
\end{align}
Using the eQP for the Higgs, see \eqref{eqphiggs}, one 
can split the function $L_S$ into a decay part, identical to the one computed in \cite{Frossard:2012pc},
\begin{align}
\label{Lrhodecay}
L^d_S( \mommaj)=16\pi{\int} &\lorentzd{\lepton \higgs}{ \momlep \momhig}   
\Ftilde{(N_i)}{\lepton \higgs}{\mommaj}{\momlep \momhig}\slashed{\momlep} \kend 
\end{align}
where we have assumed $q^0>0$, and defined 
\begin{align}
\label{defFtilde}
&\Ftilde{(a)b \dotso}{i j \dotso}{p_a p_b \dotso}{p_i p_j \dotso} \equiv 
(2\pi)^4 \delta(p_a+p_b+\dotso -p_i-p_j-\dotso) \nonumber \\
&\hspace{25mm}\times\left[f_b^{p_b}\dotso(1 \pm f_i^{p_i})(1 \pm f_j^{p_j})\right. \dotso \nonumber\\ 
&\hspace{32mm}+\left.f_i^{p_i} f_j^{p_j} \dotso(1 \pm f_b^{p_b})\dotso \right] \kend  
\end{align}
and a scattering part,
\begin{align}
\label{defLrhotop}
L_S^s&(\mommaj)=16\pi \int \lorentzdd{\momlep \momQ \momtop} (2\pi)^4 
\delta(\mommaj+\momQ-\momlep-\momtop) \nonumber \\ 
& \times \pLeptsc{}{\rho}(\momlep) \pQsc{}{\rho}(\momQ) \pTopq{}{\rho}(\momtop) 
\pHiggs{2}{R+A}(\momtop-\momQ) \EffAmplitude{}{\higgs\bar \topq\rightarrow \bar \Q}\slashed{\momlep} \nonumber \\ 
&\times \bigl[ f_\Q^\momQ (1-f_\lepton^\momlep) (1-f_\topq^\momtop)+
f_\lepton^\momlep f_\topq^\momtop (1-f_Q^\momQ)\bigr]\dend
\end{align}
Performing the frequency integration as explained above, see \eqref{lepcurSE}, we can 
rewrite \eqref{defLrhotop} as a sum of four terms, corresponding to the three scattering 
and one three-body decay process as well as their \CP-conjugates. Assuming $q^0>0$ we obtain:
\begin{align}
\label{lrhosimple}
L_S^s(\mommaj)&=16 \pi \int \! \dpi{\lepton}{ \Q \topq }{\momlep}{ \momQ \momtop}\nonumber\\
&\times \bigl[ \Ftilde{(N_i)\Q}{\lepton \topq}{\mommaj \momQ}{\momlep \momtop} 
\pHiggs{2}{R+A}(\momtop-\momQ)\EffAmplitude{}{\higgs\Q\rightarrow \topq}
\nonumber\\
&+\Ftilde{(N_i) \bar \topq}{\lepton \bar \Q}{\mommaj  \momtop}{\momlep \momQ} 
\pHiggs{2}{R+A}(\momtop-\momQ)\EffAmplitude{}{\higgs\bar \topq\rightarrow \bar \Q} 
\nonumber \\
&+\Ftilde{(N_i)\bar \lepton}{\bar \Q \topq}{\mommaj \momlep}{\momQ \momtop} 
\pHiggs{2}{R+A}(\momtop+\momQ)\EffAmplitude{}{\higgs\rightarrow\bar \Q \topq} \nonumber\\
&+\Ftilde{(N_i)}{\lepton \bar \Q \topq}{\mommaj }{\momlep \momQ \momtop} 
\pHiggs{2}{R+A}(\momtop+\momQ)\EffAmplitude{}{\higgs\Q\rightarrow \topq}  \bigr]\slashed{\momlep}\dend
\end{align}
In the regime $m_\higgs<m_\Q+m_\topq$ the intermediate Higgs cannot be on-shell. Therefore one 
can neglect the Higgs width in $\pHiggs{2}{R+A}$ and approximate it  by $\pHiggs{2}{T}$. 

As can be inferred from the definition \eqref{defFtilde}  for the scattering terms $\tilde {\cal F}$ 
vanishes in vacuum, whereas for the decay term it does not. Due to Lorentz covariance 
in vacuum both $L_S^d$ and $L_S^s$ must be proportional to the four-vector $q$. Using 
\eqref{ellphiintegral} and \eqref{eqn: decay integration limits} we find that the coefficient 
of proportionality is equal to unity for the decay contribution, i.e. $L_S^d=q$, if thermal masses 
of the Higgs and leptons are neglected. Using \eqref{NtoLQtReactDens} we find that for the scattering
contribution the coefficient of proportionality reads: 
\begin{align}
\label{Lsvacuum}
\frac{g_s \yuqSqu}{16\pi^2}& \int_{(m_\Q+m_\topq)^2}^{(M_i-m_\lepton)^2}
\frac{ds}{M_i^2}\lambda^\frac12(1,x_\Q,x_\topq)\lambda^\frac12(1,x_\lepton,x_\higgs)\nonumber\\
&\times \frac{(s-m_\topq^2-m_\Q^2)(M_i^2+m_\lepton^2-s)}{(s-m_\higgs^2)^2}\dend
\end{align}
Note that since we have omitted the Higgs decay width, this expression is convergent only 
if $m_\higgs< m_\Q+m_\topq$. The vacuum result \eqref{Lsvacuum} provides also a very good approximation for 
nonzero temperatures provided that $M/T\gg 1$. The thermal masses of the quarks then ensure 
that the Higgs remains off-shell and therefore that \eqref{Lsvacuum} is finite. It is 
important to note that due to the temperature dependence of the effective masses the 
coefficient \eqref{Lsvacuum} is temperature-dependent as well. A numerical analysis shows 
that it grows as the temperature decreases. 

\begin{figure}[h!]
 \includegraphics[width=0.95\columnwidth]{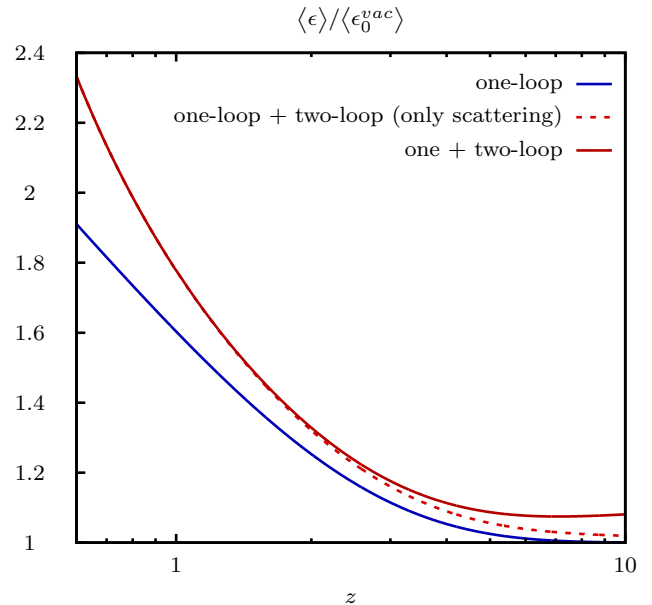}
\caption{\label{NLH_CP}Ratio of the thermally averaged \CP-violating parameter to the one calculated
in vacuum neglecting the contribution of the Higgs-mediated processes. The blue line corresponds to 
\eqref{Lrhodecay}, whereas the red lines to the sum of \eqref{Lrhodecay} and \eqref{lrhosimple}.
The dashed red line is obtained by omitting the contribution of the three-body decay in \eqref{lrhosimple}.}
\end{figure}

Using $L_S$ we can calculate the in-medium \CP-violating parameter in $\majneutrino_i \leftrightarrow 
\lepton \higgs$ process. For a hierarchical mass spectrum \cite{Frossard:2012pc}:
\begin{align}
\epsilon=\epsilon^{vac}_0\frac{\momlep L_S}{\mommaj \momlep}\kend
\end{align}
where $\epsilon^{vac}_0$ denotes the vacuum \CP-violating parameter calculated neglecting contributions 
of the Higgs-mediated processes, i.e. using only $L_S^d$. As has been mentioned above, if thermal masses 
of the Higgs and leptons are neglected then $L_S^d=q$ in vacuum and we recover $\epsilon=\epsilon_0^{vac}$. 
Once the contribution of the Higgs-mediated processes is taken into account $\epsilon^{vac}\neq \epsilon^{vac}_0$.
To estimate the size of the corrections induced by \eqref{lrhosimple} we plot the ratio of thermally 
averaged \CP-violating parameter, $\langle \epsilon\rangle\equiv \langle \epsilon \gamma^D_N \rangle
/ \langle \gamma^D_N \rangle$, to $\epsilon_0^{vac}$, see \fig\ref{NLH_CP}. 
Note that we have neglected thermal masses of the final-state Higgs and lepton in the numerics. The blue 
line corresponds to the \CP-violating parameter computed using \eqref{Lrhodecay}. In agreement with the above discussion 
the ratio reaches unity at low temperatures. The dashed red line  corresponds to the \CP-violating parameter 
computed using the sum of \eqref{Lrhodecay} and the \textit{scattering} (lines two to four) contributions 
to \eqref{lrhosimple}. As expected, at high temperatures we observe an enhancement of the ratio, whereas 
at low temperatures it reaches unity. The solid red line is obtained by considering the sum of \eqref{Lrhodecay} 
and all of the terms in \eqref{lrhosimple}. Since the three-body process is kinematically suppressed at high
temperatures, the dashed an solid lines overlap for $z\lesssim 1$. At lower temperatures the quantum-statistical 
effects are small. However, in agreement with the discussion below \eqref{Lsvacuum}, the effective thermal
masses of the Higgs and quarks lead to a slow rise of the ratio at low temperatures.

\end{appendix}

% =============================================================================
%\bibliographystyle{apsrev}
%\bibliography{../references}

% =============================================================================

\end{document}